\documentclass[pra,twocolumn,superscriptaddress]{revtex4-1}

\usepackage[utf8]{inputenc}
\usepackage[english]{babel}
\usepackage[T1]{fontenc}

\usepackage{graphicx}
\usepackage{amsmath}
\usepackage{amsfonts}
\usepackage{amssymb}

\usepackage{tikz-cd}

\usepackage{dsfont}
\usepackage{microtype}

\usepackage[breaklinks=true,colorlinks]{hyperref}

\newcommand{\cf}{cf.\ }

\newcommand{\coloneq}{\mathrel{\mathop:}=}
\newcommand{\eqcolon}{=\mathrel{\mathop:}}
\newcommand{\dd}{\mathrm{d}}
\newcommand{\Tr}{\operatorname{Tr}}
\newcommand{\ket}[1]{\left|{#1}\right\rangle}

\newcommand{\ketbra}[2]{\left|{#1}\middle\rangle\middle\langle{#2}\right|}
\newcommand{\proj}[1]{\ketbra{#1}{#1}}

\newcommand{\kB}{k_\mathrm{B}}
\newcommand{\nbar}{\bar{n}}
\newcommand{\Srel}[2]{S\left({#1}\middle\|{#2}\right)}

\begin{document}

\title{Quantum engine efficiency bound beyond the second law of thermodynamics}

\author{Wolfgang Niedenzu}
\email{Wolfgang.Niedenzu@weizmann.ac.il}
\affiliation{Department of Chemical Physics, Weizmann Institute of Science, Rehovot~7610001, Israel}

\author{Victor Mukherjee}
\affiliation{Department of Chemical Physics, Weizmann Institute of Science, Rehovot~7610001, Israel}
\affiliation{Department of Physics, Shanghai University, Baoshan District, Shanghai~200444, P.\,R.~China}

\author{Arnab Ghosh}
\affiliation{Department of Physics, Shanghai University, Baoshan District, Shanghai~200444, P.\,R.~China}
\affiliation{Department of Chemical Physics, Weizmann Institute of Science, Rehovot~7610001, Israel}

\author{Abraham G. Kofman}
\affiliation{Department of Chemical Physics, Weizmann Institute of Science, Rehovot~7610001, Israel}
\affiliation{CEMS, RIKEN, Saitama, 351-0198, Japan}

\author{Gershon Kurizki}
\affiliation{Department of Chemical Physics, Weizmann Institute of Science, Rehovot~7610001, Israel}

\begin{abstract}
  According to the second law, the efficiency of cyclic heat engines is limited by the Carnot bound that is attained by engines that operate between two thermal baths under the reversibility condition whereby the total entropy does not increase. Quantum engines operating between a thermal and a squeezed-thermal bath have been shown to surpass this bound. Yet, their maximum efficiency cannot be determined by the reversibility condition, which may yield an unachievable efficiency bound above unity. Here we identify the fraction of the exchanged energy between a quantum system and a bath that necessarily causes an entropy change and derive an inequality for this change. This inequality reveals an efficiency bound for quantum engines energised by a non-thermal bath. This bound does not imply reversibility, unless the two baths are thermal. It cannot be solely deduced from the laws of thermodynamics.
\end{abstract}

\date{October 29, 2017}

\maketitle

\section{Introduction}

Engines are machines that convert some form of energy (e.g., thermal or electrical energy) into work. Their efficiency, defined as the ratio of the extracted work to the invested energy, is restricted to $1$ at most by the energy-conservation law. While mechanical engines may reach this bound, Carnot showed~\cite{carnotbook} that the efficiency of any heat engine that cyclically operates between two thermal baths is universally limited by the ratio of the bath temperatures, regardless of the concrete design~\cite{schwablbook,kondepudibook}. The universality of this bound led to the introduction of the notion of entropy by Clausius~\cite{clausius1865verschiedene} and the formalisation of the second law of thermodynamics.

\par

The Carnot bound is attained by (idealised) heat engines that operate reversibly between two (cold and hot) thermal baths, so that the total entropy of the engine and the two baths combined is unaltered over a cycle~\cite{kondepudibook,schwablbook,callenbook}. This corresponds to the minimum amount of heat being dumped into the cold bath, so as to close the cycle, and hence to the maximum input heat being transformed into work. By contrast, in an irreversible cycle, a larger amount of heat must be dumped into the cold bath, so that less input heat is available for conversion into work, causing the engine efficiency to decrease~\cite{kondepudibook,callenbook}. 

\par

Whereas the above considerations hold for engines that operate between two thermal baths at temperatures $T_\mathrm{c}$ and $T_\mathrm{h}$, there are more general engine cycles that comprise additional baths at intermediate temperatures between $T_\mathrm{c}$ and $T_\mathrm{h}$. However, any such cycle (be it reversible or not) is less efficient than a reversible cycle that solely involves $T_\mathrm{c}$ and $T_\mathrm{h}$~\cite{schwablbook}. Hence, to find out how to use available resources most efficiently it suffices to consider the two-bath scenario.

\par

As part of the effort to understand the rapport between quantum mechanics and thermodynamics~\cite{scovil1959three,pusz1978passive,lenard1978thermodynamical,alicki1979quantum,scully2003extracting,allahverdyan2004maximal,erez2008thermodynamic,delrio2011thermodynamic,horodecki2013fundamental,correa2014quantum,skrzypczyk2014work,brandao2015second,pekola2015towards,uzdin2015equivalence,campisi2016power,rossnagel2016single} (see~\cite{kosloff2013quantum,gelbwaser2015thermodynamics,goold2016role,vinjanampathy2016quantum,kosloff2017quantum} for recent reviews), the Carnot bound has been challenged for quantum engines in which one or both of the baths are non-thermal~\cite{scully2003extracting,dillenschneider2009energetics,huang2012effects,abah2014efficiency,rossnagel2014nanoscale,hardal2015superradiant,niedenzu2016operation,manzano2016entropy,klaers2017squeezed,agarwalla2017quantum}. In this respect, a distinction is to be drawn between two types of non-thermal engines~\cite{niedenzu2016operation,dag2016multiatom}, (i) engines wherein the working medium equilibrates to a thermal state whose temperature is adjustable (e.g., by the phase of the coherence in a ``phaseonium'' bath~\cite{scully2003extracting}), which qualify as genuine heat engines with a controllable Carnot bound, and (ii) engines wherein the non-thermal (e.g., squeezed~\cite{rossnagel2014nanoscale}) bath may render the working-medium state non-thermal, making the Carnot bound irrelevant.

\par

The efficiency bound of the latter type of engines has been addressed~\cite{abah2014efficiency,rossnagel2014nanoscale,niedenzu2016operation,manzano2016entropy,agarwalla2017quantum} but still needs elucidation. What is particularly puzzling is that, contrary to heat engines that operate between two thermal baths, their efficiency bound cannot be deduced from the requirement of reversible operation: Reversibility may entail an efficiency bound that not only surpasses the (as mentioned, irrelevant) Carnot bound but also unity~\cite{manzano2016entropy}, making it unachievable. Hence, the question naturally arises whether such engines are limited by constraints other than the second law.

\par

The second law for quantum relaxation processes is widely accepted~\cite{alicki1979quantum,alicki2004thermodynamics,boukobza2007three,parrondo2009entropy,deffner2011nonequilibrium,boukobza2013breaking,kosloff2013quantum,sagawa2013second,argentieri2014violation,binder2015quantum,gelbwaser2015thermodynamics,uzdin2015equivalence,goold2016role,manzano2016entropy,vinjanampathy2016quantum,brandner2016periodic,breuerbook} to be faithfully rendered by Spohn's inequality~\cite{spohn1978entropy}. According to this inequality, the entropy change of a system that interacts with a thermal bath is bounded from below by the exchanged energy divided by the bath temperature. What has not been considered so far is, however, that the bound on entropy change in quantum relaxation processes crucially depends on whether the state of the relaxing system is non-passive. The definition~\cite{pusz1978passive,lenard1978thermodynamical,allahverdyan2004maximal} of a non-passive state~\cite{pusz1978passive,lenard1978thermodynamical,allahverdyan2004maximal,anders2013thermodynamics,alicki2013entanglement,gelbwaser2013work,hovhannisyan2013entanglement,binder2015quantacell,binder2015quantum,gelbwaser2015thermodynamics,perarnau2015extractable,skrzypczyk2015passivity,brown2016passivity,dag2016multiatom,depalma2016passive,goold2016role,niedenzu2016operation,vinjanampathy2016quantum,bruschi2017gravitational} is that its energy can be unitarily reduced until the state becomes passive, thereby extracting work. Non-passive states may thus be thought of as being ``quantum batteries''~\cite{alicki2013entanglement,binder2015quantacell} or ``quantum flywheels''~\cite{levy2016quantum}. The maximum amount of work extractable from such states (their ``work capacity'') has been dubbed ``ergotropy'' in Ref.~\cite{allahverdyan2004maximal}. For example, every population-inverted state is non-passive and so are, e.g., coherent or squeezed field states, whereas thermal states are passive.

\par

Here we examine the adequacy of assessing the maximum efficiency via the standard reversibility criterion in experimentally-relevant~\cite{rossnagel2016single,klaers2017squeezed} cyclic engines that intermittently interact with two (thermal or non-thermal) baths. We show that the standard reversibility criterion provides an inequality for the change in the engine entropy which may be much too loose (non-tight) to be useful if non-passive states are involved. The distinction between non-passive and passive states is at the heart of our analysis and underlies our division of the energy exchanged between a quantum system and a bath into a part that necessarily causes an entropy change, and ergotropy. Our proposed division is in fact a new unraveling of the first law of thermodynamics for quantum systems. In scenarios where non-thermal baths may create non-passive states of the working medium, we derive a new inequality for the entropy change which yields a physical efficiency limit of the engine that never surpasses unity. This efficiency limit in general cannot be assessed by the standard reversibility criterion. We illustrate these results for the practically-relevant Carnot- and Otto cycles~\cite{klaers2017squeezed} energised by non-thermal baths. Both cycles are shown to be restricted by our new efficiency bound.

\section{The first law of quantum thermodynamics}

For an arbitrary process taking the initial state $\rho_0$ of a quantum system to an evolving state $\rho(t)$, which may be governed by a time-dependent Hamiltonian $H(t)$ and a bath, energy conservation implies
\begin{equation}\label{eq_first_law}
  \Delta E(t)=\mathcal{E}_\mathrm{d}(t)+W(t),
\end{equation}
where $\Delta E(t)$ is the change in the system energy $E(t)=\Tr[\rho(t) H(t)]$. Its two constituents are 
\begin{subequations}\label{eq_defs_Ediss_work}
  \begin{equation}\label{eq_def_DeltaEdiss}
    \mathcal{E}_\mathrm{d}(t)\coloneq\int_0^t\Tr[\dot\rho(t^\prime)H(t^\prime)]\dd t^\prime,
  \end{equation}
  which is the non-unitary dissipative energy change due to the interaction with the bath, and
  \begin{equation}\label{eq_def_work}
    W(t)\coloneq\int_0^t\Tr[\rho(t^\prime)\dot H(t^\prime)]\dd t^\prime,
  \end{equation}
\end{subequations}
which is the work~\cite{pusz1978passive} due to changes of the system Hamiltonian. Contrary to the energy change $\Delta E(t)$, both $\mathcal{E}_\mathrm{d}(t)$ and $W(t)$ are process variables that generally depend on the evolution path, not only on the initial and final states. For thermal baths, the energy~\eqref{eq_def_DeltaEdiss} is commonly identified with the transferred heat~\cite{alicki1979quantum}. The energy $\mathcal{E}_\mathrm{d}(t)$ vanishes for a closed (isolated) system whose state evolves unitarily according to the von~Neumann equation $\dot\rho(t)=\frac{1}{i\hbar}[H(t),\rho(t)]$. The work~\eqref{eq_def_work} is either extracted or invested by the external agent that controls the system via a time-dependent Hamiltonian, as in driven engines.

\par

We here consider general scenarios, wherein the bath and/or the system may be in a non-thermal state and strive to better understand the nature of the exchanged energy~\eqref{eq_def_DeltaEdiss} and, in particular, its relation to entropy change. As we show, only part of the exchanged energy $\mathcal{E}_\mathrm{d}(t)$ is necessarily accompanied by a change in entropy.

\par

\begin{figure}
  \centering
  \includegraphics[width=0.85\columnwidth]{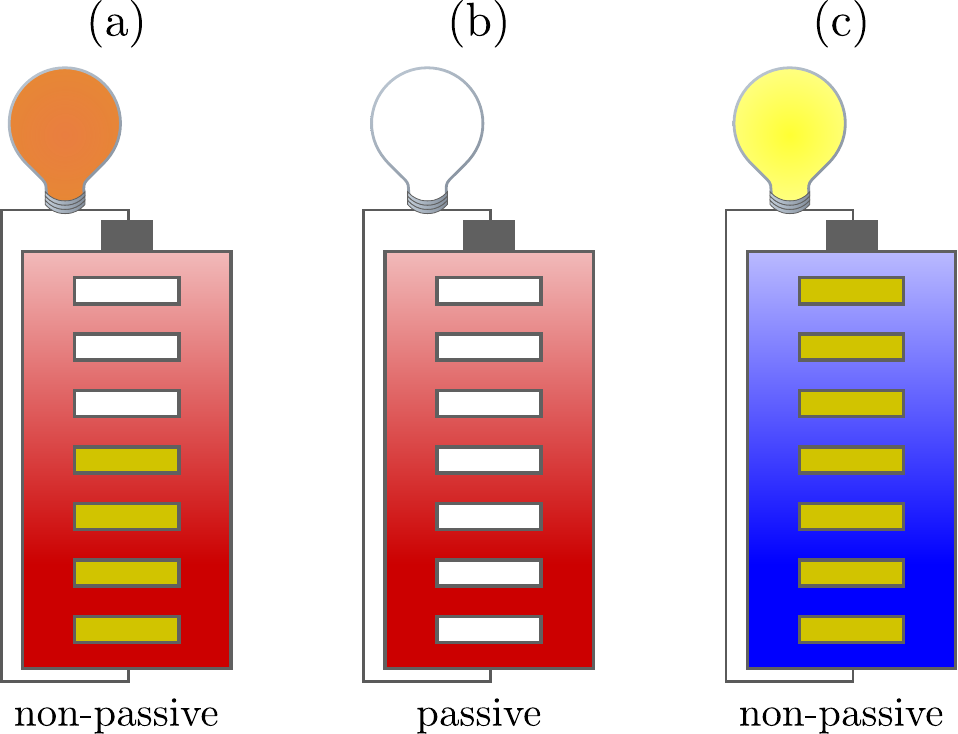}
  \caption{\textbf{Visualisation of the concept of passive energy and ergotropy.} The different kinds of energy contained in a quantum state visualised by means of a battery at a certain temperature. The battery charge (yellow bars) represents ergotropy~$\mathcal{W}$ (extractable as work, here illustrated by a lighted bulb) and its temperature (colour of the battery: red---hot, blue---cold) represents passive (here: thermal) energy $E_\mathrm{pas}$---the higher the temperature the larger the passive energy. (a) The battery is partly charged and hot: This represents a non-passive state which allows for work extraction. As the battery is not completely charged, the light bulb appears dim. (b) The battery is discharged, but its temperature is the same as in (a). This state is the passive state of (a) and, consequently, the light bulb does not shine. (c) The battery is in a non-passive state whose ergotropy is higher than in (a) (the battery is fully charged) but the passive energy is lower (the battery is colder). Although the total energy in (a) and (c) may be the same, more work can be extracted from the state~(c), causing the light bulb to shine brighter than in~(a).}\label{fig_ergotropy}
\end{figure}

\par

To elucidate this issue, we resort to the concept of non-passive states (see Fig.~\ref{fig_ergotropy} and Appendix~\ref{app_ergotropy}). The energy $E(t)$ of a non-passive state $\rho(t)$ can be decomposed into ergotropy $\mathcal{W}(t)\geq0$ and passive energy $E_\mathrm{pas}(t)$. Ergotropy is the maximum amount of work that can be extracted from such a state by means of unitary transformations~\cite{pusz1978passive,lenard1978thermodynamical,allahverdyan2004maximal}. By contrast, the passive energy, which is the energy of the passive state $\pi(t)$, cannot be extracted in the form of work.

\par

The von~Neumann entropy $\mathcal{S}(\rho(t))=-\kB\Tr[\rho(t)\ln\rho(t)]$ of a non-passive state $\rho(t)$ is the same as that of its passive state $\pi(t)$ since the two are related by a unitary transformation. Hence, a change in entropy requires a change in the passive state $\pi(t)$. Equation~\eqref{eq_def_DeltaEdiss}, however, does not discriminate between $\rho(t)$ and $\pi(t)$: A change in $\rho(t)$ may cause a non-zero $\mathcal{E}_\mathrm{d}(t)$ but not necessarily a change in entropy. By contrast, a change in $\pi(t)$ results in entropy change.

\par

In order to explicitly account for a change in the passive state, we may decompose the dissipative energy change~\eqref{eq_def_DeltaEdiss} as follows,
\begin{equation}\label{eq_DeltaEdiss_decomposition}
  \mathcal{E}_\mathrm{d}(t)=\Delta E_\mathrm{pas}|_\mathrm{d}(t)+\Delta\mathcal{W}|_\mathrm{d}(t),
\end{equation}
where
\begin{subequations}\label{eq_defs_heat_passive_work}
  \begin{equation}\label{eq_def_heat}
    \Delta E_\mathrm{pas}|_\mathrm{d}(t)\coloneq\int_{0}^{t} \Tr[\dot{\pi}(t^\prime)H(t^\prime)]\dd t^\prime
  \end{equation}
  is the dissipative (non-unitary) change in passive energy and
  \begin{equation}\label{eq_def_DeltaW_diss}
    \Delta\mathcal{W}|_\mathrm{d}(t)\coloneq\int_0^t\Tr\Big[\big(\dot\rho(t^\prime)-\dot\pi(t^\prime)\big)H(t^\prime)\Big]\dd t^\prime
  \end{equation}
\end{subequations}
is the dissipative (non-unitary) change in the system ergotropy due to its interaction with the bath. The microscopic decomposition of the exchanged energy~\eqref{eq_DeltaEdiss_decomposition} into dissipative change in passive energy~\eqref{eq_def_heat} and dissipative ergotropy change~\eqref{eq_def_DeltaW_diss} is a new unraveling of the first law of thermodynamics for quantum systems that constitutes one of our main results. 

\par

The decomposition~\eqref{eq_DeltaEdiss_decomposition} carries with it the following insights: (a)~Although ergotropy may be transferred from a non-thermal bath to the system in a non-unitary fashion, it may afterwards still be extracted from the system in the form of work via a suitable unitary transformation. (b)~Consistently, any unitary changes (in either ergotropy or in passive energy due to time-dependent changes of the Hamiltonian) are associated with work~\eqref{eq_def_work}. If the Hamiltonian is constant, then $\Delta E_\mathrm{pas}|_\mathrm{d}(t)$ is only the change in passive energy without work, $\Delta E_\mathrm{pas}|_\mathrm{d}(t)=\Delta E_\mathrm{pas}(t)=\Tr[\pi(t)H]-\Tr[\pi_0H]$, where $\pi_0$ is the passive counterpart of the initial state $\rho_0$. Likewise, $\Delta\mathcal{W}|_\mathrm{d}(t)=\Delta\mathcal{W}(t)=\mathcal{W}(\rho(t))-\mathcal{W}(\rho_0)$ is then the change in ergotropy without work performance. (c)~While a non-zero $\Delta E_\mathrm{pas}|_\mathrm{d}(t)$ entails a change in the passive state $\pi(t)$ and hence in entropy, a non-zero $\mathcal{E}_\mathrm{d}(t)$, by contrast, does not necessarily imply an entropy change, as shown below. The correspondence of $\Delta E_\mathrm{pas}|_\mathrm{d}(t)$ and $\Delta\mathcal{S}(t)$ is plausible since they have the same sign provided a majorisation relation~\cite{mari2014quantum,binder2015quantum} holds for $\rho(t)$, as detailed in Appendix~\ref{app_majorisation}.

\par

\begin{figure}
  \centering
  \includegraphics[width=0.90\columnwidth]{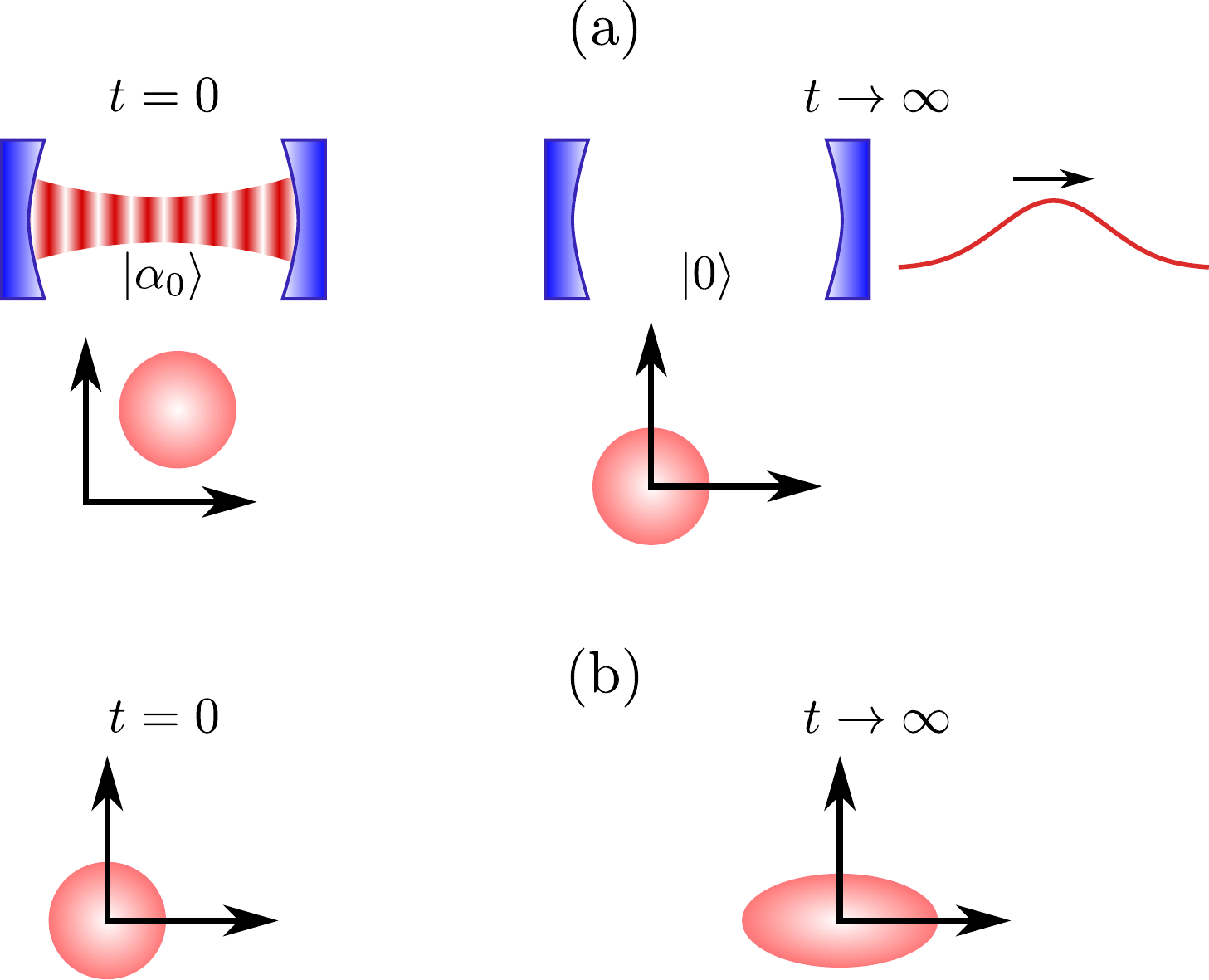}
  \caption{\textbf{Interaction of a cavity mode with thermal and non-thermal baths.} (a)~A cavity mode initialised in a coherent state decays into the surrounding electromagnetic-field bath to the vacuum state. (b)~A cavity mode prepared in the vacuum state evolves to a squeezed-vacuum state due to its interaction with a squeezed bath. The circles and the ellipse represent the respective phase-space distributions~\cite{gardinerbook} of the field states.}\label{fig_examples}
\end{figure}

\par

\par

Let us illustrate these insights for a single cavity mode (harmonic oscillator at frequency $\omega$) prepared in a pure coherent state $\rho_0=\proj{\alpha_0}$ that interacts (via a leaky mirror) with the surrounding electromagnetic-field bath (Fig.~\ref{fig_examples}a), which for optical frequencies is very close to the vacuum state~\cite{gardinerbook}. Being in contact with a bath, the cavity-mode state evolves in a non-unitary fashion (according to a quantum master equation~\cite{breuerbook}). Since the Hamiltonian is constant, the work~\eqref{eq_def_work} vanishes, $W(t)=0$. While the cavity field exponentially decays to the vacuum state, $\rho(t)=\proj{\alpha_0e^{-i\omega t-\kappa t}}$, where $\kappa$ is the leakage rate, its entropy does not change, $\mathcal{S}(\rho(t))=0$, so that the passive state $\pi(t)=\proj{0}$ is constant. Consequently, $\Delta E_\mathrm{pas}|_\mathrm{d}(t)=0$ and the entire energy change is due to dissipated ergotropy, $\Delta E(t)=\Delta\mathcal{W}|_\mathrm{d}(t)=\hbar\omega|\alpha_0|^2(e^{-2\kappa t}-1)\leq0$.

\par

\begin{figure}
  \centering
  \includegraphics[width=0.90\columnwidth]{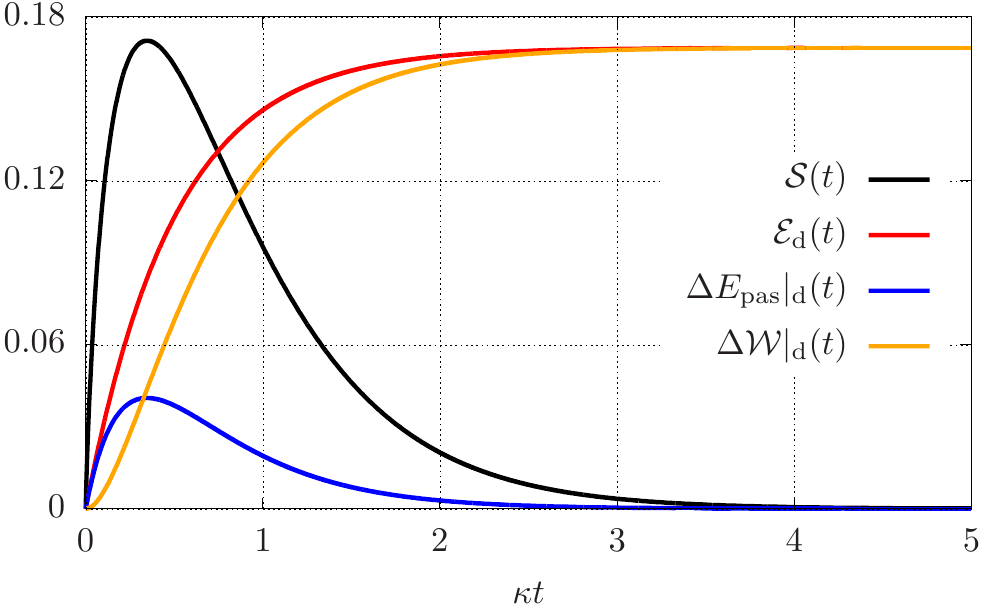}
  \caption{\textbf{Entropy and energy of a cavity mode interacting with a squeezed bath.} Entropy, ergotropy and energy changes for a single cavity mode prepared in the vacuum state that interacts with an outside bath in a squeezed-vacuum state (Fig.~\ref{fig_examples}b) obtained by a numerical integration of the master equation. The energies are given in units of $\hbar\omega$ and the entropy in units of $\kB$. Parameters: $\omega=10\kappa$ and squeezing parameter $r=0.4$, $\kappa$ being the decay rate of the cavity.}\label{fig_squeezed_cavity}
\end{figure}

\par

As another example, consider again a single cavity mode, this time prepared in its vacuum state $\rho_0=\proj{0}$, that interacts with an outside bath in a squeezed-vacuum state~\cite{gardinerbook} (Appendix~\ref{app_master_equation_squeezed_bath}), eventually converging to a squeezed-vacuum state inside the cavity (Fig.~\ref{fig_examples}b). Although the initial and the steady state have zero entropy, this is not true during the evolution (Fig.~\ref{fig_squeezed_cavity}). Consequently, both dissipative passive-energy change $\Delta E_\mathrm{pas}|_\mathrm{d}(t)$ and dissipative ergotropy change $\Delta\mathcal{W}|_\mathrm{d}(t)\geq 0$ occur. Figuratively, this process corresponds to a non-unitary charging of a battery.

\section{Reversibility criterion}\label{sec_spohn}

In non-equilibrium thermodynamics, the accepted criterion for the irreversibility or reversibility of the system relaxation to its steady state is the non-negativity of the entropy production~\cite{kondepudibook}. For quantum systems that are weakly coupled to (thermal or non-thermal) Markovian baths, Spohn~\cite{spohn1978entropy} put forward an expression for the entropy production $\Sigma(t)$. Here, we are interested in relaxation to steady state, for which we define $\Sigma\coloneq\Sigma(\infty)$, satisfying (Appendix~\ref{app_spohn})
\begin{equation}\label{eq_spohn_integrated}
  \Sigma\geq0,
\end{equation}
where the equality sign is the reversibility condition. For a constant Hamiltonian, it evaluates to $\Sigma=\Srel{\rho_0}{\rho_\mathrm{ss}}\geq0$, where $\Srel{\rho_0}{\rho_\mathrm{ss}}\coloneq\kB\Tr[\rho_0(\ln \rho_0-\ln \rho_\mathrm{ss})]$ is the entropy of the system initialised in a state $\rho_0$ at $t=0$ relative to the steady state $\rho_\mathrm{ss}$ to which it relaxes. For a slowly time-varying Hamiltonian~\cite{alicki1979quantum,alipour2016correlations}, Eq.~\eqref{eq_spohn_integrated} gives rise to an inequality for the the change $\Delta\mathcal{S}$ of the system (von~Neumann) entropy, given in Appendix~\ref{app_spohn}.

\par

The common~\cite{alicki1979quantum,alicki2004thermodynamics,boukobza2007three,parrondo2009entropy,deffner2011nonequilibrium,boukobza2013breaking,kosloff2013quantum,sagawa2013second,argentieri2014violation,binder2015quantum,gelbwaser2015thermodynamics,uzdin2015equivalence,brandner2016periodic,goold2016role,manzano2016entropy,vinjanampathy2016quantum,breuerbook} identification of Eq.~\eqref{eq_spohn_integrated} with the second law appears plausible for systems in contact with thermal baths: It then evaluates to $\Sigma=\Delta\mathcal{S}-\mathcal{E}_\mathrm{d}/T\geq0$, where $\mathcal{E}_\mathrm{d}$ is the dissipative change in the system energy defined in Eq.~\eqref{eq_def_DeltaEdiss} (in the limit $t\rightarrow\infty$).

\par

Here we contend that although inequality~\eqref{eq_spohn_integrated} is a formally correct statement of the second law (under standard thermodynamic assumptions), it may not provide a meaningful estimate of $\Delta\mathcal{S}$ if a system is initialised in a non-passive state and/or interacts with a non-thermal bath. Physically, this is because, as discussed above, the exchanged energy $\mathcal{E}_\mathrm{d}$ may be non-zero even if the entropy does not change.

\section{Entropy change in relaxation processes involving ergotropy}\label{sec_decay}

Consider the decay of an initially non-passive state $\rho_0$ to a (passive) thermal state $\rho_\mathrm{th}$ via contact with a thermal bath at temperature $T$. Based on the decomposition~\eqref{eq_DeltaEdiss_decomposition}, the reversibility condition~\eqref{eq_spohn_integrated} evaluates to (at $t\rightarrow\infty$)
\begin{equation}\label{eq_DeltaS_QdT_DeltaW}
  \Delta\mathcal{S}\geq\frac{\mathcal{E}_\mathrm{d}}{T}=\frac{\Delta E_\mathrm{pas}|_\mathrm{d}+\left.\Delta\mathcal{W}\right|_\mathrm{d}}{T},
\end{equation}
where both dissipative change in passive energy~\eqref{eq_def_heat} and dissipated ergotropy~\eqref{eq_def_DeltaW_diss} appear. In what follows we shall revise this inequality, which may greatly overestimate the actual entropy change. As shown below, a tight inequality for $\Delta\mathcal{S}$ is indispensable for correctly assessing the maximum efficiency of an engine.

\subsection{Constant Hamiltonian}

We first consider the case of a constant Hamiltonian. As we have seen, dissipative ergotropy change is not necessarily linked to a change in entropy. Therefore, the lower bound on $\Delta\mathcal{S}$ in Eq.~\eqref{eq_DeltaS_QdT_DeltaW} may be not tight (maximal). It is obtained from Spohn's inequality~\eqref{eq_spohn_integrated} for the relaxation of an initially non-passive state in a thermal bath. However, one may resort to the fact that the entropy $\mathcal{S}$ is a state variable, so that $\Delta\mathcal{S}=\mathcal{S}(\rho_\mathrm{th})-\mathcal{S}(\rho_0)$ is path-independent, i.e., its value only depends on the initial state $\rho_0$ and the (passive) thermal steady state $\rho_\mathrm{th}$. Hence, Spohn's inequality~\eqref{eq_spohn_integrated} may well be applied to alternative evolution paths from $\rho_0$ to $\rho_\mathrm{th}$, giving rise to different inequalities for the same $\Delta\mathcal{S}$.

\par

In particular, we now consider a path that does not involve any dissipation of ergotropy to the bath: Namely, one may start the process by performing a unitary transformation to the passive state, $\rho_0\mapsto\pi_0$. Thereafter, this state is brought in contact with the thermal bath, yielding the steady-state solution $\rho_\mathrm{th}$. Inequality~\eqref{eq_spohn_integrated} applied to this alternative path yields
\begin{equation}\label{eq_DeltaS_QdT}
  \Delta\mathcal{S}\geq\frac{\Delta E_\mathrm{pas}|_\mathrm{d}}{T},
\end{equation}
where $\Delta E_\mathrm{pas}|_\mathrm{d}$ is the same as in Eq.~\eqref{eq_DeltaS_QdT_DeltaW}.

\par

The steady state attained via contact with a thermal bath is passive, hence the system ergotropy must decrease as a result of the relaxation, $\Delta\mathcal{W}|_\mathrm{d}=-\mathcal{W}_0\leq0$, where $\mathcal{W}_0\geq 0$ is the initial ergotropy stored in the state $\rho_0$. Hence, inequality~\eqref{eq_DeltaS_QdT} always entails inequality~\eqref{eq_DeltaS_QdT_DeltaW} and is thus a tighter and more relevant estimate of $\Delta\mathcal{S}$. This has a crucial consequence: If the initial state is non-passive, inequality~\eqref{eq_DeltaS_QdT} rules out the equality sign in inequality~\eqref{eq_DeltaS_QdT_DeltaW}, so that the considered decay via contact with a thermal bath can never be reversible according to criterion~\eqref{eq_spohn_integrated}. 

\par

We now consider the more general situation wherein the system is governed by a constant Hamiltonian and interacts with an arbitrary bath (that may not be parameterised by a temperature) until it reaches the steady state $\rho_\mathrm{ss}$. In order to obtain an optimal (the tightest) inequality for the entropy change $\Delta\mathcal{S}$, we here instead of inequality~\eqref{eq_spohn_integrated} ($\Sigma$ for such a bath is given in Appendix~\ref{app_sigma_non-thermal}) propose to adopt the mathematical relation
\begin{equation}\label{eq_srel}
  \Srel{\pi_0}{\pi_\mathrm{ss}}\geq 0.
\end{equation}
As shown in Appendix~\ref{app_sigmap_optimal}, Eq.~\eqref{eq_srel} provides generally a tight inequality for $\Delta\mathcal{S}$. The motivation for Eq.~\eqref{eq_srel} is, as before, that the entropy of any state $\rho$ is the same as that of its passive counterpart $\pi$. If $\pi_\mathrm{ss}$ is a thermal state, we recover Eq.~\eqref{eq_DeltaS_QdT}.

\par

We stress that, contrary to Spohn's inequality (Appendix~\ref{app_spohn}), Eqs.~\eqref{eq_DeltaS_QdT} and~\eqref{eq_srel} do not require weak coupling between the system and the bath (in the same spirit as in Refs.~\cite{schloegl1966zur,deffner2011nonequilibrium}) and are thus universally-valid whenever the reduced state of the system reaches a steady state.

\subsection{Time-dependent Hamiltonian}

We now allow the Hamiltonian $H(t)$ to slowly vary during the evolution~\cite{alicki1979quantum}. Contrary to the case of a constant Hamiltonian, the dissipative passive-energy change~\eqref{eq_def_heat} and the ergotropy change~\eqref{eq_def_DeltaW_diss} in the r.h.s.\ of Eq.~\eqref{eq_DeltaS_QdT_DeltaW} are now path-dependent. Namely, they are not only determined by the initial state $\rho_0$ and the steady state $\rho_\mathrm{th}(\infty)$, which is a thermal state under the Hamiltonian $H(\infty)$.

\par

Since during the evolution the time-dependent Hamiltonian may generate a non-passive state (even if the initial state is passive and the bath is thermal) we cannot, in general, find an alternative path void of dissipated ergotropy for the same $H(t)$. Notwithstanding, we may still consider a path void of initial ergotropy in the spirit of the previous section by extracting the ergotropy of the initial state in a unitary fashion prior to the interaction with the bath, resulting in the passive state $\pi_0$. Afterwards, this passive state is brought into contact with the thermal bath, yielding the steady state $\rho_\mathrm{th}(\infty)$. Spohn's inequality can be applied to the latter step, yielding
\begin{equation}\label{eq_DeltaS_Qth}
  \Delta\mathcal{S}\geq\frac{\mathcal{E}_\mathrm{d}^\prime}{T},
\end{equation}
with the energy
\begin{equation}\label{eq_def_Qth}
  \mathcal{E}_\mathrm{d}^\prime\coloneq\int_0^\infty\Tr[\dot\varrho(t)H(t)]\dd t
\end{equation}
exchanged with the bath along the alternative path. Here $\varrho(t)$ is the solution of the same thermal master equation that governs $\rho(t)$ but with the initial condition $\varrho_0=\pi_0$. In the case that the initial state $\rho_0$ is already passive, we have $\varrho(t)=\rho(t)$, $\mathcal{E}_\mathrm{d}^\prime=\mathcal{E}_\mathrm{d}$ and Eqs.~\eqref{eq_DeltaS_Qth} and~\eqref{eq_spohn_integrated} coincide. For a constant Hamiltonian, Eq.~\eqref{eq_DeltaS_Qth} evaluates to Eq.~\eqref{eq_DeltaS_QdT}.

\par

Consider now the more general situation where a quantum system interacts with a non-thermal bath and eventually relaxes to a unitarily-transformed thermal state $U\rho_\mathrm{th}(\infty)U^\dagger$. A prime example is a harmonic oscillator that interacts with a squeezed thermal bath~\cite{gardinerbook,ekert1990canonical}: Its steady state is a squeezed thermal state. Then one can show (Appendix~\ref{app_liouvillian}) that this situation can be traced back to the interaction of a unitarily-transformed state $\tilde\rho(t)\coloneq U^\dagger \rho(t) U$ with a thermal bath, provided that the Hamiltonian $H(t)$ commutes with itself at all times; a harmonic oscillator with a time-dependent frequency and time-independent eigenstates is an example. This requirement will be adopted in the remainder of this paper for any interaction of a system with a non-thermal bath. The relaxation of a possibly non-passive state $\tilde\rho(t)$ in a thermal bath pertains to the scenario considered above upon replacing $\rho(t)$ by $\tilde\rho(t)$ there. Equation~\eqref{eq_DeltaS_Qth} thus also holds for this class of non-thermal baths (the derivation and the generalisation to arbitrary non-thermal baths are discussed in Appendix~\ref{app_time-dependent_Hamiltonian}).

\par

The new entropic inequality~\eqref{eq_DeltaS_Qth} is the second main result of our work. For the special case of a constant Hamiltonian, it reduces to inequality~\eqref{eq_DeltaS_QdT}.

\section{Maximal efficiency of engines powered by non-thermal baths}\label{sec_efficiency}

In view of our new inequality~\eqref{eq_DeltaS_Qth}, does inequality~\eqref{eq_spohn_integrated} always provide a true bound on the engine efficiency? Namely, is reversibility indeed the key to operating a quantum engine at the highest possible efficiency? This question arises for cyclic engines fuelled by non-thermal (e.g., squeezed) baths, since such baths may transfer both passive thermal energy and ergotropy to the system while Eq.~\eqref{eq_spohn_integrated} does not distinguish between these two different kinds of energies.

\par

\begin{figure}
  \centering
  \includegraphics[width=0.85\columnwidth]{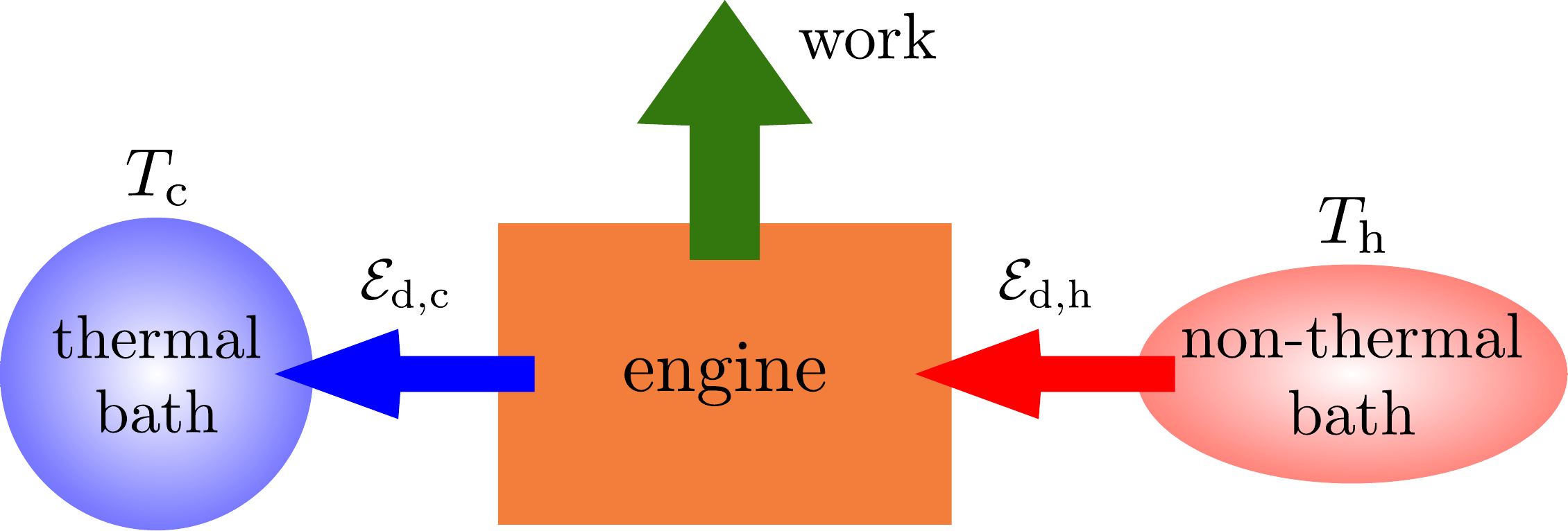}
  \caption{\textbf{Engine fuelled by a non-thermal bath.} Schematics of an engine fuelled by a hot non-thermal (e.g., squeezed thermal) bath that provides the input energy $\mathcal{E}_\mathrm{d,h}$. The engine operates in an arbitrary cycle wherein work is extracted by a piston and an amount of energy $\mathcal{E}_\mathrm{d,c}$ is dumped into the cold thermal bath.}\label{fig_engine}
\end{figure}

\par

Here we consider a quantum engine (Fig.~\ref{fig_engine}) that operates between a cold thermal bath (at temperature $T_\mathrm{c}$) and a hot non-thermal bath subject to a time-dependent drive (the ``piston''~\cite{alicki1979quantum}). As in common, experimentally-relevant situations~\cite{klaers2017squeezed}, the non-thermal bath drives the working medium into a non-passive state whose passive counterpart is assumed to be thermal. This allows us to maintain the notion of a ``hot'' bath with temperature $T_\mathrm{h}>T_\mathrm{c}$, where $T_\mathrm{h}$ is defined by the steady-state solution of the working medium. As an example, in the case of a single cavity mode interacting with the surrounding electromagnetic field in a squeezed-thermal state~\cite{breuerbook,gardinerbook}, the temperature $T_\mathrm{h}$ equals the thermodynamic temperature of the bath prior to its squeezing. The generalisation of the present analysis to arbitrary passive states is straightforward (Appendix~\ref{app_efficiency}).

\par

Existing treatments of engines powered by non-thermal baths have taken the system--baths interaction to be isochoric, i.e., subject to a constant Hamiltonian~\cite{huang2012effects,abah2014efficiency,rossnagel2014nanoscale,hardal2015superradiant,manzano2016entropy,niedenzu2016operation}. We here relax this restriction and allow for stroke cycles wherein the working-medium (WM) Hamiltonian is allowed to slowly change during the interaction with the baths~\cite{alicki1979quantum}. We only impose the condition that the WM attains its steady state at the end of the energising stroke (wherein it interacts with the hot non-thermal bath) and the resetting stroke (wherein it interacts with the cold thermal bath).

\par

The energising stroke is described by a master equation~\cite{breuerbook} that evolves the WM state to a unitarily-transformed thermal state $\rho_\mathrm{ss}(\infty)=U\rho_\mathrm{th}(\infty)U^\dagger$, hence Eq.~\eqref{eq_DeltaS_Qth} holds. After this stroke, the WM is in a non-passive state, whose ergotropy is subsequently extracted by the piston via a suitable unitary transformation. Since we seek the efficiency bound, we assume that no ergotropy is dissipated in the cold bath (and thus lost), hence the requirement to extract it from the WM before its interaction with that bath. We note that in cycles where both baths are simultaneously coupled to the WM (as in continuous cycles~\cite{gelbwaser2013minimal}), part of the ergotropy is inevitably dissipated into the cold bath, so that such cycles are inherently less efficient than stroke cycles adhering to the above requirement.

\par

Similarly, Hamiltonians that do not commute with themselves at different times are known to reduce the efficiency due to ``quantum friction''~\cite{kosloff2013quantum,brandner2016periodic,mukherjee2016speed,kosloff2017quantum}, whereas we are here interested in principal limitations on the efficiency. Hence, during the interaction with the non-thermal bath, the Hamiltonian is assumed to commute with itself at all times, as already mentioned in the discussion on the validity of Eq.~\eqref{eq_DeltaS_Qth} for such a bath.

\par

The engine's WM must return to its initial state after each cycle. This implies that $\Delta\mathcal{S}=0$ over a cycle, hence the importance of having a tight estimate for the entropy change within each stroke. The entropy changes in the two relevant strokes satisfy $\Delta\mathcal{S}_\mathrm{c}\geq \mathcal{E}_\mathrm{d,c}/T_\mathrm{c}$ and $\Delta\mathcal{S}_\mathrm{h}\geq\mathcal{E}^\prime_\mathrm{d,h}/T_\mathrm{h}$. Here $\mathcal{E}_\mathrm{d,c}\leq0$ is the change in the WM energy due to its interaction with the cold thermal bath and $\mathcal{E}^\prime_\mathrm{d,h}\geq0$ is the change the WM energy would have, had the non-thermal bath been thermal [as in Eq.~\eqref{eq_DeltaS_Qth}]. Taking into account that the WM is passive prior to its interaction with the cold bath, so that Eqs.~\eqref{eq_DeltaS_Qth} and~\eqref{eq_spohn_integrated} coincide for that stroke, the condition of vanishing entropy change over a cycle (which must hold in any cycle) then yields the inequality
\begin{equation}\label{eq_condition_gen}
   \Delta\mathcal{S}_\mathrm{c}+\Delta\mathcal{S}_\mathrm{h}=0\quad\Rightarrow\quad\frac{\mathcal{E}_\mathrm{d,c}}{T_\mathrm{c}}+\frac{\mathcal{E}^\prime_\mathrm{d,h}}{T_\mathrm{h}}\leq 0.
\end{equation}

\par

The efficiency of the engine is defined as the ratio of the extracted work to the invested energy, $\eta\coloneq-W/\mathcal{E}_\mathrm{d,h}$, where $\mathcal{E}_\mathrm{d,h}$ is the total energy (the sum of passive thermal energy and ergotropy) imparted by the non-thermal bath during the energising stroke. Using the first-law statement~\eqref{eq_first_law}, this ratio may be expressed through the energy transfers $\mathcal{E}_\mathrm{d,c}$ and $\mathcal{E}_\mathrm{d,h}$. Condition~\eqref{eq_condition_gen} on $\mathcal{E}_\mathrm{d,c}$ (the energy lost to the cold bath) then restricts the efficiency to
\begin{equation}\label{eq_etamax_gen}
  \eta\leq 1-\frac{T_\mathrm{c}}{T_\mathrm{h}}\frac{\mathcal{E}^\prime_\mathrm{d,h}}{\mathcal{E}_\mathrm{d,h}}\eqcolon\eta_\mathrm{max}.
\end{equation}
Its derivation as well as a more general expression for the case where the passive state after the energising stroke is non-thermal are given in Appendix~\ref{app_efficiency}.

\par

The efficiency bound~\eqref{eq_etamax_gen} does not only depend on the two temperatures, which is to be expected, as non-thermal baths may occur in various forms that cannot be universally described by a common set of parameters. The physical details of the bath (e.g., its squeezing parameter) are thus encoded in the fraction of the two energies $\mathcal{E}^\prime_\mathrm{d,h}$ and $\mathcal{E}_\mathrm{d,h}$, whose forms are universal. This fraction expresses the ratio of generalised heat transfer to the total energy input from the hot bath.

\par

The bound~\eqref{eq_etamax_gen} underscores the physicality of our inequality~\eqref{eq_DeltaS_Qth}: In the usual regime of functioning of the engine, $\mathcal{E}^\prime_\mathrm{d,h}\geq0$ and $\mathcal{E}_\mathrm{d,h}>0$ (i.e., the hot bath provides energy and increases the WM entropy), the bound~\eqref{eq_etamax_gen} is limited by unity, $\eta_\mathrm{max}\leq 1$, which is reached in the ``mechanical''-engine limit $\mathcal{E}^\prime_\mathrm{d,h}\rightarrow 0$ where the non-thermal bath only provides ergotropy. By contrast, the bound $\eta_\Sigma$ that stems from the reversibility condition~\eqref{eq_spohn_integrated} (derived in Appendix~\ref{app_efficiency}) may surpass $1$ (see Ref.~\cite{manzano2016entropy}). In the opposite, heat-engine, limit $\mathcal{E}^\prime_\mathrm{d,h}\rightarrow \mathcal{E}_\mathrm{d,h}$ where only passive thermal energy but no ergotropy is imparted by the hot bath, Eq.~\eqref{eq_etamax_gen} reproduces the Carnot bound $\eta_\mathrm{C}=1-T_\mathrm{c}/T_\mathrm{h}$. As shown below, if the Hamiltonian is kept constant during the interaction with the non-thermal bath, then Eq.~\eqref{eq_etamax_gen} is restricted by $\eta_\mathrm{C}\leq\eta_\mathrm{max}\leq\eta_\Sigma$. Therefore, for such engines our new bound~\eqref{eq_etamax_gen} is always tighter than the second-law bound $\eta_\Sigma$.

\par

The bound~\eqref{eq_etamax_gen} is valid in the regime $\mathcal{E}_\mathrm{d,c}\leq0$ and $\mathcal{E}^\prime_\mathrm{d,h}\geq 0$ wherein the cold bath serves as an energy dump. As shown in~\cite{niedenzu2016operation}, there exists a regime wherein such a machine acts simultaneously as an engine and a refrigerator for the cold bath. The efficiency then evaluates to $\eta=1$ (see Appendix~\ref{app_efficiency}).

\par

We have thus reached a central conclusion: The efficiency bound of the engine increases with the decrease of the ratio of the energy that an alternative thermal engine would have received (in the same energising stroke) to the total energy imparted by the non-thermal bath (in the actual engine cycle). In the limit of thermal baths~\cite{alicki1979quantum} we recover the standard Carnot bound for the efficiency of heat engines, even if the engine (in any cycle) exhibits quantum signatures (e.g., quantum coherence in the WM due to the piston action~\cite{uzdin2015equivalence}) or the WM--bath interactions are time-dependent~\cite{mukherjee2016speed}.

\par

We note that the costs of bath preparation or the heat generated by a clock~\cite{erker2017autonomous,woods2016autonomous} required to implement a time-periodic Hamiltonian will reduce the efficiency. In the spirit of thermodynamics, however, the bound~\eqref{eq_etamax_gen} only takes into account limitations inherent to the cycle.

\par

Whilst our analysis is focused on the two-bath situation, Eq.~\eqref{eq_condition_gen} can be generalised to cycles where the working medium intermittently interacts with additional (thermal or non-thermal) baths. This generalisation shows (Appendix~\ref{app_multibath}) that the efficiency of multi-bath engines is always lower than the maximum efficiency~\eqref{eq_etamax_gen} of the appropriate two-bath engine, thus reaffirming the generality of the bound~\eqref{eq_etamax_gen}.

\section{Specific quantum engines}

We now pose the question: Which bound is more relevant, $\eta_\Sigma$ (whose explicit form is given in  Appendix~\ref{app_efficiency}) that stems from the reversibility condition~\eqref{eq_spohn_integrated}, or $\eta_\mathrm{max}$ given by Eq.~\eqref{eq_etamax_gen}? Contrary to the Carnot bound, the efficiency bound~\eqref{eq_etamax_gen} not only depends on the parameters of the baths but also on the energising stroke through the stroke's initial condition and the Hamiltonian that determine the integrals $\mathcal{E}^\prime_\mathrm{d,h}$ and $\mathcal{E}_\mathrm{d,h}$. Yet, the functional form~\eqref{eq_etamax_gen} is independent of the choice of the non-thermal bath or the WM. Whether or not this bound is reached by an engine that implements this chosen energising stroke is then determined by condition~\eqref{eq_condition_gen}.

\par

In complete generality, the tighter of the alternative efficiency bounds derived here,
\begin{equation}\label{eq_eta_minimum}
  \eta\leq\min\{\eta_\mathrm{max},\eta_\Sigma\},
\end{equation}
is the relevant one. Relation~\eqref{eq_eta_minimum} is the universal thermodynamic limit on quantum engine efficiency, which never surpasses unity.

\par

Notwithstanding the alternatives that may be offered by Eq.~\eqref{eq_eta_minimum}, we now discuss two generic practically-relevant engine cycles for which one can explicitly show that $\eta_\mathrm{max}\leq\eta_\Sigma$. Such engines are thus not restricted by the second law, but by other constraints on their entropy.

\subsection{Time-dependent Hamiltonian: A squeezed photonic Carnot engine}

\par

\begin{figure}
  \centering
  \includegraphics[width=0.70\columnwidth]{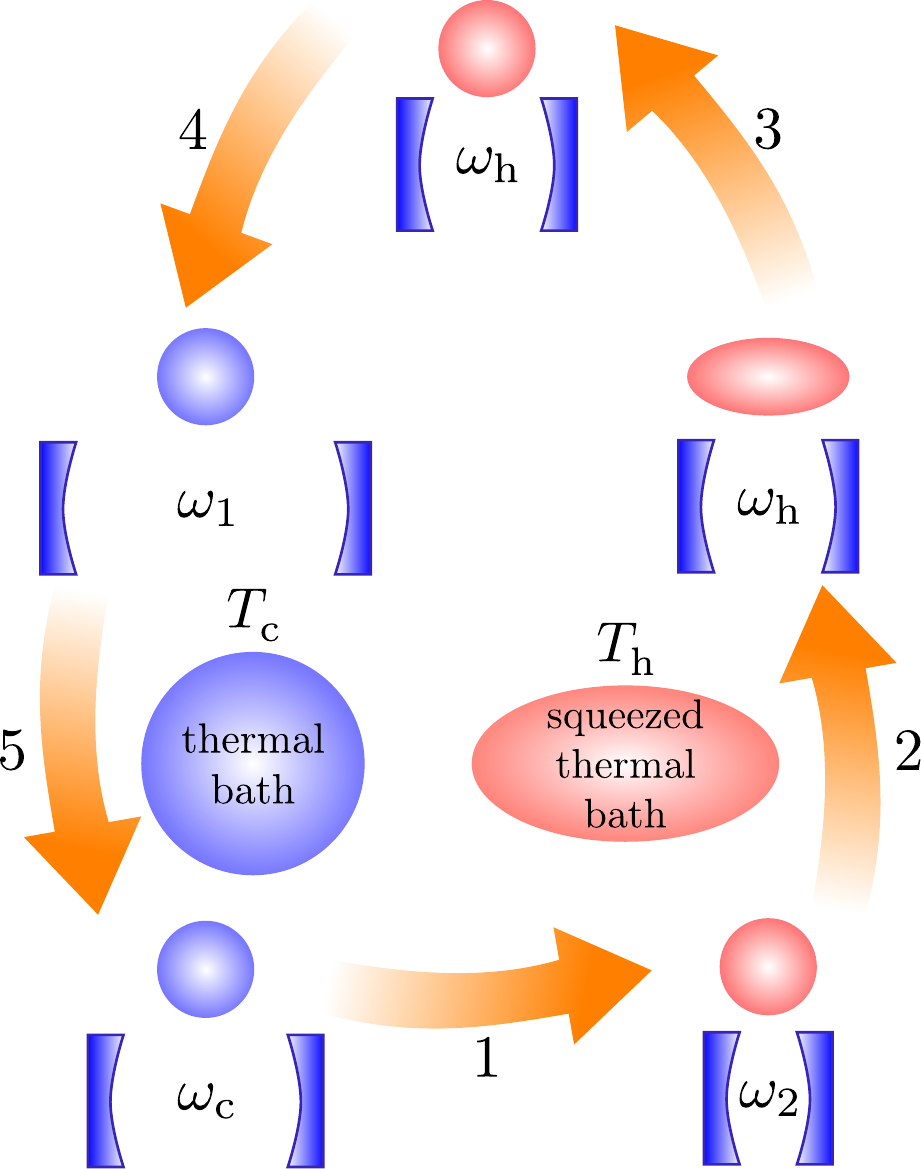}
  \caption{\textbf{A photonic Carnot cycle for a squeezed thermal bath.} The cycle starts with a thermal state with frequency $\omega_\mathrm{c}$ and temperature $T_\mathrm{c}$ (lower left corner). In stroke~$1$, the mode undergoes an adiabatic compression to frequency $\omega_2=\omega_\mathrm{c}T_\mathrm{h}/T_\mathrm{c}$ and temperature $T_\mathrm{h}>T_\mathrm{c}$. Thereafter, in the energising stroke~$2$, the frequency is slowly reduced to $\omega_\mathrm{h}\leq\omega_2$ while the mode is connected to the squeezed thermal bath, yielding a squeezed thermal steady state. Its ergotropy is extracted in stroke~$3$ by an ``unsqueezing'' unitary operation, resulting in a thermal state with temperature $T_\mathrm{h}$. In stroke~$4$, the frequency is again adiabatically reduced to $\omega_1=\omega_\mathrm{h}T_\mathrm{c}/T_\mathrm{h}$ such that the mode attains the temperature $T_\mathrm{c}$. Finally, stroke~$5$ is an isothermal compression back to the initial state.}\label{fig_carnot}
\end{figure}

\par

We first consider a photonic Carnot-like engine fuelled by a squeezed-thermal bath, as depicted in Fig.~\ref{fig_carnot}. It contains the four strokes of the regular thermal Carnot cycle~\cite{carnotbook,clausius1865verschiedene,schwablbook,kondepudibook}, as well as an additional ergotropy-extraction stroke (stroke $3$ in the figure). In the regular thermal Carnot cycle, the interactions with the baths are isothermal. 

\par

\begin{figure}
  \centering
  \includegraphics[width=0.95\columnwidth]{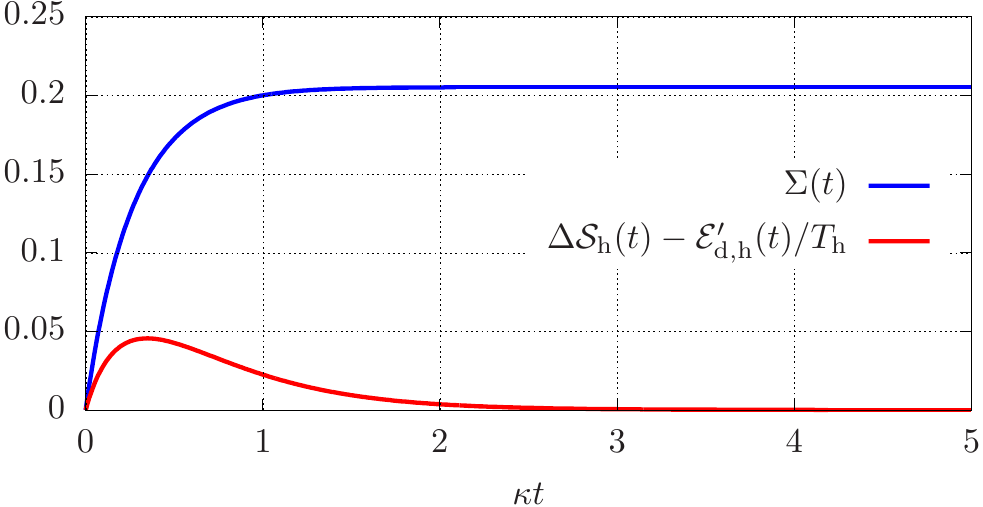}
  \caption{\textbf{Entropy change in a Carnot cycle.} Change in entropy (in units of $\kB$) during stroke~$2$ of the modified Carnot cycle in Fig.~\ref{fig_carnot} as a function of the stroke duration obtained by a numerical integration of the master equation. The upper (blue) curve corresponds to the reversibility criterion~\eqref{eq_spohn_integrated}; it is seen that the inequality $\Sigma\geq0$ is far from being saturated. By contrast, our proposed inequality~\eqref{eq_DeltaS_Qth} is saturated (i.e., the equality sign applies) for sufficiently long stroke duration (red lower curve); here $\Delta\mathcal{S}_\mathrm{h}(t)=\mathcal{S}(\rho(t))-\mathcal{S}(\rho_0)$. Parameters: Oscillator frequency $\omega(t)=(25-0.05\kappa t)\kappa$, $\kB T_\mathrm{h}=5\hbar\kappa$ and squeezing parameter $r=0.2$, $\kappa$ being the decay rate of the cavity.}\label{fig_carnot_engine}
\end{figure}

\par

Based on Eq.~\eqref{eq_DeltaS_Qth}, we have in the second stroke $\mathcal{E}^\prime_\mathrm{d,h}=T_\mathrm{h}\Delta\mathcal{S}_\mathrm{h}$, since the master equation void of squeezing induces isothermal expansion wherein the state $\varrho(t)$ is always in thermal equilibrium (Fig.~\ref{fig_carnot_engine}). Stroke $5$ is isothermal compression, i.e., $\mathcal{E}_\mathrm{d,c}=T_\mathrm{c}\Delta\mathcal{S}_\mathrm{c}$. The condition of vanishing entropy change over a cycle, $\Delta\mathcal{S}=\mathcal{E}_\mathrm{d,c}/T_\mathrm{c}+\mathcal{E}^\prime_\mathrm{d,h}/T_\mathrm{h}=0$, corresponds to the equality sign in condition~\eqref{eq_condition_gen}. Hence, the efficiency of this cycle is the bound in Eq.~\eqref{eq_etamax_gen}.

\par

Consequently, the bound $\eta_\mathrm{max}$ is lower than $\eta_\mathrm{\Sigma}$ for all possible engine cycles that contain a ``Carnot-like'' energising stroke, namely, a stroke characterised by a slowly-changing Hamiltonian and an initial thermal state at temperature $T_\mathrm{h}$, such that $\mathcal{E}^\prime_\mathrm{d,h}=T_\mathrm{h}\Delta\mathcal{S}_\mathrm{h}$.

\par

\begin{figure}
  \centering
  \includegraphics[width=0.80\columnwidth]{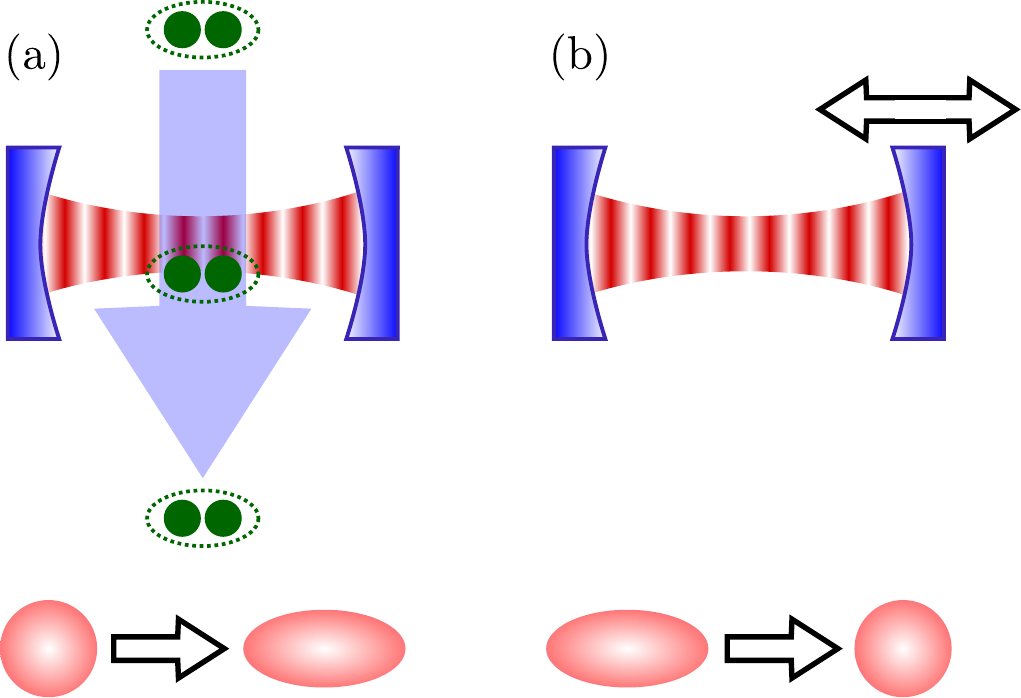}
  \caption{\textbf{Squeezing and unsqueezing of a cavity mode.} (a) The interaction of a cavity mode with a squeezed thermal bath (stroke~2 in Fig.~\ref{fig_carnot}) may be realised in a micromaser setup where a beam of entangled atom pairs passes through the cavity~\cite{dag2016multiatom}. (b) The unsqueezing operation in stroke~$3$ of Fig.~\ref{fig_carnot} may be implemented by a suitable modulation of the cavity frequency~\cite{graham1987squeezing,agarwal1991exact,averbukh1994enhanced}.}\label{fig_implementation}
\end{figure}

\par

Such a photonic Carnot engine energised by a squeezed bath may be implemented as a modification of the photonic Carnot cycle based on a cavity in a micromaser setup in the seminal work by Scully et al.~\cite{scully2003extracting}: Instead of a beam of coherently-prepared three-level atoms (``phaseonium'') that constitute an effective thermal bath for the cavity-mode WM, we here suggest, following Ref.~\cite{dag2016multiatom}, to use a beam of suitably-entangled atom pairs passing through a cavity that may act as a squeezed-thermal bath for the same WM (Fig.~\ref{fig_implementation}a). The steady state of the cavity mode is then determined by a squeezing parameter $r$ and a temperature $T_\mathrm{h}$, which are both a function of the two-atom state~\cite{dag2016multiatom}. A major advantage of this method is that it allows for very high squeezing parameters. In order to extract the ergotropy that is stored in the cavity mode after its interaction with the squeezed bath and before its interaction with the cold bath (where it would be lost), a unitary transformation that ``unsqueezes'' the cavity field must be performed, e.g., as in Refs.~\cite{graham1987squeezing,agarwal1991exact,averbukh1994enhanced}, where the cavity-mode frequency is abruptly ramped up and then gradually ramped down (Fig.~\ref{fig_implementation}b).

\subsection{Constant Hamiltonian: An Otto-like cycle}

Next, we consider a quantum Otto cycle~\cite{geva1992quantum,feldmann2004characteristics,quan2007quantum,delcampo2014more,uzdin2015equivalence,kosloff2017quantum} that consists of two isentropic strokes (adiabatic compression and decompression of the WM), two isochoric strokes (interaction with the baths at a fixed Hamiltonian) and an additional ergotropy-extraction stroke. This cycle amounts to setting $\omega_2=\omega_\mathrm{h}$ and $\omega_1=\omega_\mathrm{c}$ in Fig.~\ref{fig_carnot}.

\par

Since the Hamiltonian is now kept constant during the energising stroke, we have $\mathcal{E}^\prime_\mathrm{d,h}=\Delta E_\mathrm{pas,h}$, where $\Delta E_\mathrm{pas,h}$
is the change in passive energy during the hot stroke, and $\mathcal{E}_\mathrm{d,h}=\Delta E_\mathrm{pas,h}+\Delta \mathcal{W}_\mathrm{h}$, where $\Delta \mathcal{W}_\mathrm{h}$ is the change in ergotropy during that stroke. The efficiency of this Otto-like cycle is bounded by Eq.~\eqref{eq_etamax_gen},
\begin{equation}\label{eq_etamax_otto}
  \eta_\mathrm{max}^\mathrm{Otto}=1-\frac{T_\mathrm{c}}{T_\mathrm{h}}\frac{\Delta E_\mathrm{pas,h}}{\Delta E_\mathrm{pas,h}+\Delta \mathcal{W}_\mathrm{h}}\leq \eta_\Sigma,
\end{equation}
but this bound is only attained in the ``mechanical'' limit $\mathcal{E}^\prime_\mathrm{d,h}=\Delta E_\mathrm{pas,h}=0$, where only ergotropy is transferred from the non-thermal bath and no net entropy change occurs during the strokes. In this case the bound equals $1$, as one expects for mechanical engines. By contrast, the Carnot-like cycle always operates at maximum efficiency, even when both passive thermal energy and ergotropy are imparted by this bath.

\par

\begin{figure}
  \centering
  \includegraphics[width=0.95\columnwidth]{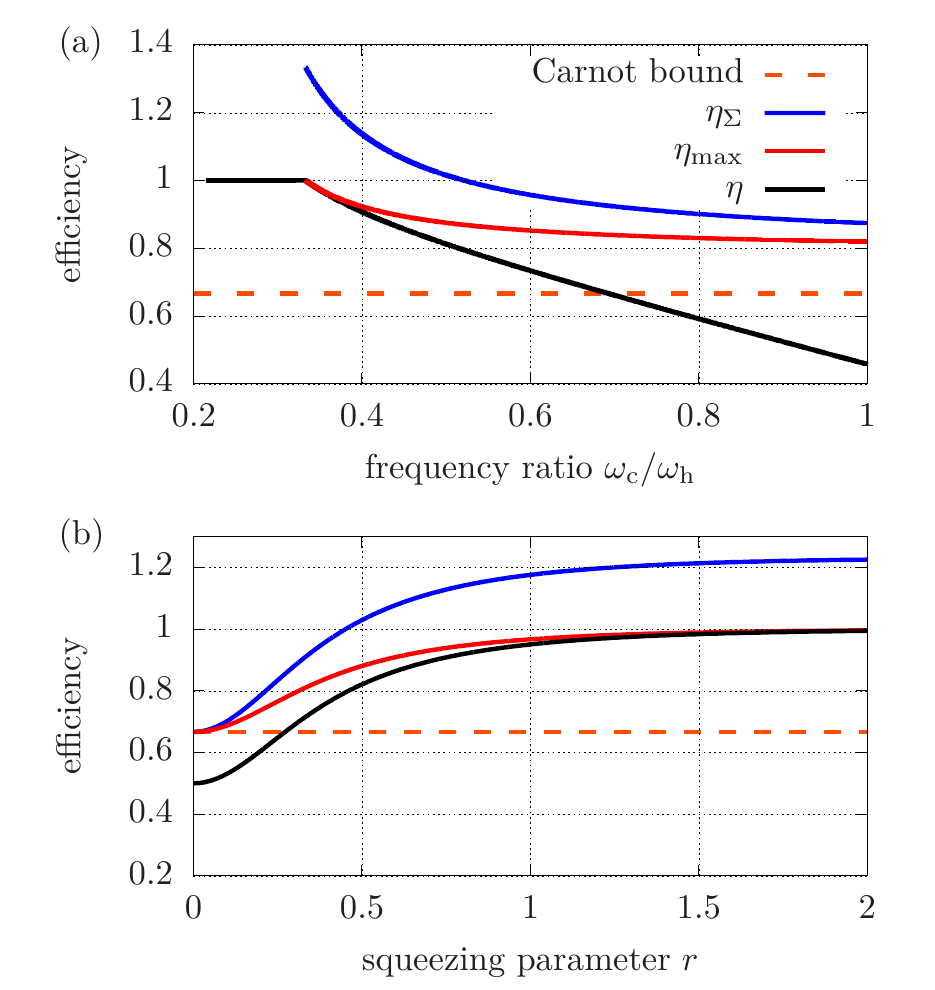}
  \caption{\textbf{Efficiency bounds for the Otto-like cycle.} Actual efficiency $\eta$ and alternative efficiency bounds (the explicit expressions are summarised in Appendix~\ref{app_figures}) for an Otto-like cycle implemented with a harmonic-oscillator working medium and a squeezed thermal bath as a function of (a) the frequency ratio and (b) the squeezing parameter. The bounds only hold in the regime $\mathcal{E}_\mathrm{d,c}\leq0$ (see text). Parameters: $T_\mathrm{h}=3T_\mathrm{c}$ and (a) squeezing parameter $r=0.5$ and (b) oscillator frequencies $\omega_\mathrm{c}/\omega_\mathrm{h}=0.5$.}\label{fig_efficiency_otto}
\end{figure}

In general, any engine cycle wherein the interaction with the hot bath is isochoric (has constant Hamiltonian) and sufficiently long (for the WM to reach steady state) abides by the bound~\eqref{eq_etamax_otto}, which is lower than the bound $\eta_\Sigma$ imposed by the second law (Fig.~\ref{fig_efficiency_otto}). Moreover, their efficiency bound always surpasses the Carnot bound, $\eta_\mathrm{max}^\mathrm{Otto}\geq\eta_\mathrm{C}$.

\par

\section{Discussion}

Our analysis has been aimed at comparing the efficiency bounds and the conditions for their attainment in quantum engines energised by thermal and non-thermal baths. These respective bounds turn out to be very different since, unlike thermal baths, non-thermal baths may exchange both thermal (passive) energy and ergotropy with the working medium (WM). To this end we have revisited the first law of thermodynamics and identified as passive energy the part of the energy exchange with the bath that necessarily causes a change in the WM entropy [Eq.~\eqref{eq_def_heat}]. This division of the exchanged energy relies on the distinction between passive and non-passive states of the WM. Only the latter states store ergotropy that may be completely extracted in the form of work. Our energetic division conceptually differs from the one involving ``housekeeping heat'' previously provided for classical systems~\cite{hatano2001steady}. It would be interesting to extend our analysis to situations where ``housekeeping heat'' has been considered in a quantum context~\cite{gardas2015thermodynamic,misra2015quantum}.

\par

Based on the distinction between passive and non-passive states, we have put forward a new estimate~\eqref{eq_DeltaS_Qth} of the entropy change in quantum relaxation processes, which turns out to be the key to understanding the limitations of quantum engines fuelled by arbitrary baths. Cyclic engines whose passive energy is altered by the baths are restricted in efficiency by limits on their entropy change. Yet, for a wide class of practically-relevant engines, including all engines whose energising stroke is either isochoric or Carnot-like, the restriction imposed by inequality~\eqref{eq_DeltaS_Qth} on the entropy change is stricter than what the second law~\eqref{eq_spohn_integrated} would allow. By contrast, the commonly used reversibility is a global condition on the WM and the two baths combined that is imposed by the second law, and hence not necessarily a relevant characterisation of engine efficiency.

\par

An alternative formulation of our main insight is that, for any baths, entropy change limits the engine efficiency in the same way as in traditional heat engines---condition~\eqref{eq_condition_gen} is the same whether the energising bath is thermal or not. Namely, maximal efficiency is reached when (a)~no ergotropy (extractable work) is dumped into the cold bath and (b)~no entropy is generated within the engine, or, equivalently, minimal energy is dumped into the cold bath~\cite{kondepudibook}. For thermal engines, this criterion of minimal energy dumping and the reversibility criterion coincide, but the two criteria differ if the energising bath is non-thermal.

\par

Another important insight is that the same efficiency bound~\eqref{eq_etamax_gen} ensues whether the WM is energised by a non-thermal bath or by a thermal bath (that supplies thermal energy) combined with a battery (that supplies ergotropy) provided the total energy imparted by the WM remains the same. This supports the description of non-thermal engines as hybrids of thermal (thermal-energy-fuelled) and ``mechanical'' (ergotropy-fuelled) engines~\cite{niedenzu2016operation}.

\par

Our theory provides better understanding of the operation principles of quantum engines: These are shown not to follow only from the laws of thermodynamics, but require discrimination between different (passive and non-passive) quantum states of the system (WM) and the baths involved. The present generalisation of the treatment of standard thermal processes for quantum systems is not only the key to the construction of the most efficient hybrid engines that are unrestricted by the Carnot bound, as in the recent experimental implementation of an engine powered by a squeezed bath~\cite{klaers2017squeezed}. It may also open a new perspective on quantum-channel communications~\cite{mari2014quantum,depalma2016passive,qi2016thermal} where entropic constraints play a major role.

\section*{Acknowledgements}
We thank the ISF, AERI and VATAT for support.

\section*{Author contributions}
W.\,N. conceived the idea. W.\,N. and A.G.\,K. performed the calculations. W.\,N. implemented and performed the numerical simulations. W.\,N., V.\,M., A.\,G, A.G.\,K. and G.\,K. contributed to the discussion and the interpretation of the results. All authors were involved in the discussion during the writing of the manuscript. W.\,N., A.G.\,K. and G.\,K. wrote the manuscript.

\appendix

\section{Non-passive states}\label{app_ergotropy}

The energy $E$ of a state $\rho$ with respect to a Hamiltonian $H$ can be decomposed into ergotropy $\mathcal{W}$ and passive energy $E_\mathrm{pas}$. Ergotropy is the maximum amount of work that can be extracted from the state by means of unitary transformations such that the Hamiltonian before and after the unitary coincide~\cite{pusz1978passive,lenard1978thermodynamical,allahverdyan2004maximal}. The passive energy, by contrast, cannot be extracted in the form of work. States that only contain passive energy are called passive states.

\par

Ergotropy is defined as
\begin{equation}\label{eq_app_def_ergotropy}
  \mathcal{W}(\rho,H)\coloneq\Tr(\rho H)-\min_U\Tr(U\rho U^\dagger H)\geq0,
\end{equation}
where the minimisation is over the set of all possible unitary transformations. Consequently, any state $\rho$ can be written as $\rho=V_\rho\pi V_\rho^\dagger$, i.e., as a unitarily-transformed passive state $\pi$, where $V_\rho$ is the unitary that realises the minimum appearing on the r.h.s.\ of Eq.~\eqref{eq_app_def_ergotropy}. The energy of the state $\rho$ thus reads
\begin{equation}
  E=E_\mathrm{pas}+\mathcal{W}=\Tr[\pi H]+\Tr[(\rho-\pi)H].
\end{equation}
Explicitly, the passive state and its energy read
\begin{subequations}
  \begin{align}
    \pi&\coloneq\sum_n r_n\proj{n}\label{eq_app_passive_state}\\
    E_\mathrm{pas}&=\Tr[\pi H]=\sum_n r_n E_n,\label{eq_app_passive_energy}
  \end{align}
\end{subequations}
where $\{r_n\}$ are the ordered ($r_{n+1}\leq r_n\,\forall n$) eigenvalues of $\rho$ and $\{\ket{n}\}$ is the ordered ($E_{n+1}\geq E_n\,\forall n$) eigenbasis of $H$. When $H$ is non-degenerate, $\pi$ is unique. If $H$ is degenerate, its eigenbasis and, consequently, the passive state~\eqref{eq_app_passive_state}, may be not unique. However, the energies~\eqref{eq_app_passive_energy} of all passive states corresponding to $\rho$ are the same and equal the passive energy of $\rho$.

\section{Majorisation relation}\label{app_majorisation}

Assume $\rho(t^\prime)\succ\rho(t^{\prime\prime})$ for any $t^{\prime\prime}\geq t^\prime$ in some time interval $I$ ($t^\prime,t^{\prime\prime}\in I$), namely that $\rho(t^\prime)$ majorises~\cite{binder2015quantum,mari2014quantum} $\rho(t^{\prime\prime})$ in this interval, i.e., 
\begin{equation}\label{eq_app_majorisation}
  \sum_{m=1}^nr_m(t^\prime)\geq\sum_{m=1}^nr_m(t^{\prime\prime})\quad (1\leq n\leq N),
\end{equation}
where $r_{m+1}(\tau)\leq r_m(\tau)$ ($\tau\in I$) are the ordered eigenvalues of $\rho(\tau)$ [\cf Eq.~\eqref{eq_app_passive_state}] and $N$ is the dimension of the Hilbert space of the system.
\par

Let us consider the sign of the dissipative passive-energy change $\Delta E_\mathrm{pas}|_\mathrm{d}$ under this majorisation condition. We may write~\eqref{eq_def_heat} in the form
\begin{equation}
  \Delta E_\mathrm{pas}|_\mathrm{d}(t) = \int_0^t\dd\tau \Tr[\dot\pi(\tau)H(\tau)]=\int_{0}^{t}\dd \tau \lim_{h\rightarrow 0}f(\tau,h),
\end{equation}
where we have defined
\begin{equation}
  f(\tau,h)\coloneq\sum_{n=1}^N \frac{r_n(\tau+h)-r_n(\tau)}{h} E_n(\tau),
\end{equation}
where $E_{n+1}(\tau)\geq E_n(\tau)$ are the ordered eigenvalues of the Hamiltonian [\cf Eq.~\eqref{eq_app_passive_energy}]. Using summation by parts and the normalisation of the density matrix, this function may be rewritten as
\begin{equation}
  f(\tau,h)=\sum_{n=1}^{N-1}[E_{n+1}(\tau)-E_n(\tau)]\sum_{m=1}^n\frac{r_m(\tau)-r_m(\tau+h)}{h}.
\end{equation}
The first factor is non-negative due to the monotonically-ordered energies. The second factor is also non-negative if Eq.~\eqref{eq_app_majorisation} holds in the entire integration domain $[0,t]$. In this case, the majorisation relation implies $\Delta E_\mathrm{pas}|_\mathrm{d}(t)\geq 0$.

\par

Let us now turn to the sign of the entropy change. If $\rho_1\succ\rho_2$, then $\mathcal{S}(\rho_2)\geq\mathcal{S}(\rho_1)$~\cite{allahverdyan2004maximal}. Hence, we have the relation
\begin{multline}\label{eq_majorisation_Q_DeltaS_succ}
  \rho(t^\prime)\succ\rho(t^{\prime\prime})\quad\forall\ 0\leq t^\prime\leq t^{\prime\prime}\leq t\\\Rightarrow\quad \Delta E_\mathrm{pas}|_\mathrm{d}(t)\geq 0\,\wedge\, \Delta\mathcal{S}(t)\geq 0,
\end{multline}
where $\Delta\mathcal{S}(t)=\mathcal{S}(\rho(t))-\mathcal{S}(\rho_0)$. Similarly, one can show that the opposite relation holds, $\rho(t^\prime)\prec\rho(t^{\prime\prime})\Rightarrow \Delta E_\mathrm{pas}|_\mathrm{d}(t)\leq 0\,\wedge\, \Delta\mathcal{S}(t)\leq 0$. When the Hamiltonian is non-degenerate, $\Delta E_\mathrm{pas}|_\mathrm{d}(t)$ and $\Delta\mathcal{S}(t)$ can be shown to vanish iff the passive state corresponding to $\rho(\tau)$ is constant (i.e., the evolution of $\rho(\tau)$ is unitary) for $\tau \in [0,t]$.

\par

For the case of a constant Hamiltonian, relation~\eqref{eq_majorisation_Q_DeltaS_succ} was obtained in Ref.~\cite{binder2015quantum}. In this case, $\Delta E_\mathrm{pas}|_\mathrm{d}(t)=\Delta E_\mathrm{pas}(t)$ and hence Eq.~\eqref{eq_majorisation_Q_DeltaS_succ} implies that the passive energy of $\rho_2$ is greater than or equal to the passive energy of $\rho_1$ if $\rho_1\succ\rho_2$ or, equivalently, if $\pi_1\succ\pi_2$, where $\pi_i$ is the passive state corresponding to $\rho_i$ ($i=1,2$).

\section{Master equation for a squeezed bath}\label{app_master_equation_squeezed_bath}

In the interaction picture, the master equation for a harmonic oscillator that interacts with a squeezed thermal bath reads~\cite{gardinerbook}
\begin{multline}\label{eq_app_master_squeezing}
  \dot\rho=\kappa(N+1)\mathcal{D}(a,a^\dagger)[\rho]+\kappa N\mathcal{D}(a^\dagger,a)[\rho]\\-\kappa M\mathcal{D}(a,a)[\rho]-\kappa M\mathcal{D}(a^\dagger,a^\dagger)[\rho],
\end{multline}
where $\mathcal{D}(A,B)[\rho]\coloneq 2A\rho B-BA\rho-\rho BA$. Here $\kappa$ denotes the decay rate and (w.l.o.g.\ we have set the squeezing phase to zero)
\begin{subequations}
  \begin{align}\label{eq_master_squeezing_standard_coefficients}
    N&\coloneq\bar{n}(\cosh^2r+\sinh^2r)+\sinh^2r\\
    M&\coloneq-\cosh r\sinh r (2\bar{n}+1),
  \end{align}
\end{subequations}
where $\bar{n}=[\exp(\hbar\omega/[\kB T])-1]^{-1}$ is the thermal excitation number of the bath at the oscillator frequency $\omega$ and $r$ the squeezing parameter. The results in Fig.~\ref{fig_squeezed_cavity} were obtained by a numerical solution of Eq.~\eqref{eq_app_master_squeezing} with $\bar{n}=0$.

\par

Defining $b\coloneq S(r)a S^\dagger(r)=a\cosh r+a^\dagger\sinh r$, where $S(r)=\exp\left[\frac{r}{2}a^2-\frac{r}{2}(a^\dagger)^2\right]$ is the unitary squeezing operator, the master equation~\eqref{eq_app_master_squeezing} can be cast into the Lindblad form~\cite{breuerbook,ekert1990canonical}
\begin{equation}
  \dot\rho=\kappa(\bar{n}+1)\mathcal{D}(b,b^\dagger)[\rho]+\kappa \bar{n}\mathcal{D}(b^\dagger,b)[\rho].
\end{equation}
Its steady-state solution is the squeezed thermal state $S(r)\left[Z^{-1}\exp\left(-\hbar\omega a^\dagger a/[\kB T]\right)\right]S^\dagger(r)$.

\section{Entropy production $\Sigma$}\label{app_spohn}

Spohn's inequality for the entropy-production rate reads~\cite{spohn1978entropy}
\begin{equation}\label{eq_spohn}
  \sigma\coloneq-\frac{\dd}{\dd t}\Srel{\rho(t)}{\rho_\mathrm{ss}}\geq0,
\end{equation}
where $\Srel{\rho(t)}{\rho_\mathrm{ss}}\coloneq\kB\Tr[\rho(t)(\ln \rho(t)-\ln \rho_\mathrm{ss})]$. Inequality~\eqref{eq_spohn} holds for any $\rho(t)$ that evolves according to a Lindblad master equation~\cite{breuerbook}
\begin{equation}\label{eq_app_master}
  \dot\rho=\mathcal{L}\rho,
\end{equation}
$\mathcal{L}$ being the Liouvillian (Lindblad operator). The steady-state solution of Eq.~\eqref{eq_app_master} obeys $\mathcal{L}\rho_\mathrm{ss}=0$. Then, upon defining $\Sigma\coloneq\int_0^{\infty}\sigma\dd t$, the time-integrated inequality~\eqref{eq_spohn} yields
\begin{equation}\label{eq_spohn_integrated_app}
   \Sigma=\Srel{\rho_0}{\rho_\mathrm{ss}}\geq0.
\end{equation}

\par

Equality~\eqref{eq_spohn} requires the coupling between the system and the bath to be sufficiently weak and the bath relaxation to be sufficiently fast to allow for the perturbative derivation of the Lindblad master equation. In the spirit of traditional thermodynamics, the Lindblad approach excludes correlations or entanglement between the system and the bath~\cite{breuerbook}. In general, Eq.~\eqref{eq_spohn} may not hold for non-Markovian baths~\cite{erez2008thermodynamic}. In contrast, since the relative entropy is non-negative, Eq.~\eqref{eq_spohn_integrated_app} holds for arbitrary coupling between the system and the bath~\cite{schloegl1966zur,deffner2011nonequilibrium}.

\par

As shown in Refs.~\cite{alicki1979quantum,alipour2016correlations}, Spohn's inequality~\eqref{eq_spohn} can be generalised to time-dependent Hamiltonians under the condition that $H(t)$ varies slowly compared to the relaxation time of the reservoir~\cite{alicki1979quantum}. The corresponding master equation then reads
\begin{equation}
  \dot\rho(t)=\mathcal{L}(t)\rho(t),
\end{equation}
where $\mathcal{L}(t)$ is the same Liouvillian as in Eq.~\eqref{eq_app_master}, but with time-dependent coefficients (\cf Ref.~\cite{alicki1979quantum}). Its invariant state $\rho_\mathrm{ss}(t)$ satisfies $\mathcal{L}(t)\rho_\mathrm{ss}(t)=0$. The generalisation of inequality~\eqref{eq_spohn} then reads~\cite{alipour2016correlations}
\begin{equation}\label{eq_spohn_L_t_diff}
  \sigma=\left.-\frac{\dd}{\dd s}\Srel{e^{s\mathcal{L}(t)}\rho(t)}{\rho_\mathrm{ss}(t)}\right|_{s=0}\geq0.
\end{equation}
Upon integration, Eq.~\eqref{eq_spohn_L_t_diff} evaluates to the inequality
\begin{equation}\label{eq_spohn_L_t}
  \Sigma=\Delta\mathcal{S}+\kB\int_0^\infty\Tr\Big[\big(\mathcal{L}(t)\rho(t)\big)\ln\rho_\mathrm{ss}(t)\Big]\dd t\geq0
\end{equation}
for the entropy change $\Delta\mathcal{S}=\mathcal{S}(\rho_\mathrm{ss}(\infty))-\mathcal{S}(\rho_0)$. In the case of a constant Hamiltonian, Eq.~\eqref{eq_spohn_L_t} reduces to Eq.~\eqref{eq_spohn_integrated_app}.

\par

If the Liouvillian describes the interaction with a thermal bath at temperature $T$, i.e., $\mathcal{L}(t)=\mathcal{L}_\mathrm{th}(t)$, then $\rho_\mathrm{ss}(t)=\rho_\mathrm{th}(t)$, where 
\begin{equation}\label{eq_app_rhoth_t}
  \rho_\mathrm{th}(t)=\frac{1}{Z(t)}\exp\left(-\frac{H(t)}{\kB T}\right)
\end{equation}
is a thermal state for the (instantaneous) Hamiltonian $H(t)$. Equation~\eqref{eq_spohn_L_t} then yields
\begin{equation}\label{eq_spohn_t_integrated}
  \Delta\mathcal{S}\geq\frac{1}{T}\int_{0}^{\infty}\Tr\left[\dot{\rho}(t)H(t)\right]\dd t= \frac{\mathcal{E}_\mathrm{d}}{T},
\end{equation}
with the dissipated energy $\mathcal{E}_\mathrm{d}$ defined in Eq.~\eqref{eq_def_DeltaEdiss}.

\section{Entropy production $\Sigma$ for non-thermal baths}\label{app_sigma_non-thermal}

Let us consider $\Sigma$ in the case of a constant Hamiltonian [Eq.~\eqref{eq_spohn_integrated_app}] for a non-thermal bath that gives rise to a non-passive steady state $\rho_\mathrm{ss}=U\pi_\mathrm{ss}U^\dagger$ via the Liouvillian $\mathcal{L}_U$. This $\Sigma$ can be related to that of a passive state, as follows. Since the relative entropy is invariant with respect to a unitary transformation of its arguments, Eq.~\eqref{eq_spohn_integrated_app} can be recast in the form
\begin{equation}\label{eq_sigma_nonthermal_bath}
  \Sigma=\Srel{\tilde\rho_0}{\pi_\mathrm{ss}}\geq0,
\end{equation}
where $\tilde\rho_0\coloneq U^\dagger\rho_0U$. Thus, $\Sigma$ equals the entropy production obtained under the relaxation of an open system from the unitarily-transformed state $\tilde\rho_0$ to the passive state $\pi_\mathrm{ss}$.

\par

In particular, when $\pi_\mathrm{ss}$ is the thermal state $\rho_\mathrm{th}$, $\Sigma$ equals the entropy production obtained under thermalisation of the system starting from the state $\tilde\rho_0$ and we have
\begin{equation}\label{eq_sigma_rhotilde_thermal_bath}
  \Sigma=\Delta\mathcal{S}-\frac{\tilde{\mathcal{E}}_\mathrm{d}}{T}\geq0,
\end{equation}
where $\tilde{\mathcal{E}}_\mathrm{d}$ is the change in the energy $\tilde E=\Tr[\tilde\rho H]$ of the transformed state $\tilde\rho$.

\par

Consider now a slowly-varying $H(t)$ such that inequality~\eqref{eq_spohn_L_t} holds. The invariant state of $\mathcal{L}_U(t)$ now reads $\rho_\mathrm{ss}(t)=U\rho_\mathrm{th}(t)U^\dagger$, with the (instantaneous) thermal state~\eqref{eq_app_rhoth_t}. Inequality~\eqref{eq_spohn_L_t} then yields
\begin{equation}\label{eq_sigma_t_LU_integrated}
  \Sigma=\Delta\mathcal{S}-\frac{1}{T}\int_{0}^{\infty}\Tr\left[U^\dagger\dot{\rho}(t)UH(t)\right]\dd t\geq 0,
\end{equation}
where the appearing integral is the generalisation of $\tilde{\mathcal{E}}_\mathrm{d}$ from inequality~\eqref{eq_sigma_rhotilde_thermal_bath}. It is shown in Appendix~\ref{app_liouvillian} that $U^\dagger\dot{\rho}(t)U$ equals a thermal Liouvillian acting on a unitarily-transformed state [Eq.~\eqref{eq_liouvillian_U_th_t}]. Hence, also for a time-dependent Hamiltonian, the evaluation of $\Sigma$ in a non-thermal bath reduces to the case of a transformed state that decays via contact with a thermal bath.

\section{Optimality of the inequality for relative entropy}\label{app_sigmap_optimal}

Equation~\eqref{eq_srel} provides a generally tighter inequality for $\Delta\mathcal{S}$ than Eq.~\eqref{eq_sigma_nonthermal_bath} [or~\eqref{eq_spohn_integrated_app}]. Indeed, Eq.~\eqref{eq_sigma_nonthermal_bath} can be written as $\Delta\mathcal{S}\geq\mathcal{S}(\pi_\mathrm{ss})-\kB A$, where $A=-\Tr[\tilde\rho_0\ln\pi_\mathrm{ss}]$. This inequality is the tightest (i.e., its r.h.s.\ is maximal) on the set of all states $\tilde\rho_0$ which differ from $\rho_0$ by a unitary transformation, when $A$ is minimal on this set. Note that $\pi_\mathrm{ss}$ commutes with the Hamiltonian and the eigenvalues of $-\ln\pi_\mathrm{ss}$ do not decrease as a function of the eigenvalues of the Hamiltonian. Thus, $-\ln\pi_\mathrm{ss}$ can be considered, in a sense, as an effective ``Hamiltonian'', for which $A$ is the average ``energy'' in the state $\tilde\rho_0$. The average energy amongst unitarily-accessible states is known to be minimised in the passive state. When $H$ is non-degenerate, then the passive state $\pi_0$ corresponding to $H$ is also the passive state corresponding to the effective ``Hamiltonian'' $-\ln\pi_\mathrm{ss}$; hence, $A$ is minimal for $\tilde\rho_0=\pi_0$.

\par

By contrast, if $H$ is degenerate there is generally no unique passive state (Appendix~\ref{app_ergotropy}). In this case, $A$ is minimal not for each $\pi_0$ but iff $\pi_0$ is also a passive state of the effective ``Hamiltonian'', i.e., iff $\pi_0$ commutes with $\pi_\mathrm{ss}$. One can show that there exists, at least, one such state $\pi_0$. Thus, Eq.~\eqref{eq_srel} provides the tightest inequality for $\Delta\mathcal{S}$ among all inequalities of the form~\eqref{eq_sigma_nonthermal_bath} or~\eqref{eq_spohn_integrated_app}.

\section{Unitary equivalence of non-thermal and thermal baths}\label{app_liouvillian}

The time evolution of an initial state $\rho_0$ under the Liouvillian $\mathcal{L}_U$ as defined in Appendix~\ref{app_sigma_non-thermal} may be replaced by an alternative time evolution involving a thermal bath. These two equivalent evolution paths can be lucidly represented by the diagram (see also~\cite{ekert1990canonical} and Appendix~\ref{app_master_equation_squeezed_bath})
\begin{equation}\label{eq_cd}
  \begin{tikzcd}
    \rho_0 \arrow[r,"\mathcal{L}_U",mapsto] \arrow[d,"U^\dagger" left,mapsto,bend right,dashed] & \rho(t)=U\tilde\rho(t)U^\dagger \\ \tilde{\rho}_0=U^\dagger\rho_0 U \arrow[r,"\mathcal{L}_\mathrm{th}",mapsto,bend right,dashed] & \tilde\rho(t) \arrow[u,"U" right,mapsto,bend right,dashed]
  \end{tikzcd}.
\end{equation}
According to Eq.~\eqref{eq_cd}, the evolution of $\rho_0$ induced by a non-thermal bath towards $\rho_\mathrm{ss}$ (solid arrow) may be replaced by a three-stage process (dashed arrows) wherein the system is in contact with a thermal bath only in the second step.

\par

This may be shown as follows. The Liouvillian $\mathcal{L}_U$ in the interaction picture may be cast into the general Lindblad form~\cite{breuerbook}
\begin{equation}\label{eq_LU_Lalpha}
  \mathcal{L}_U\rho=\sum_\alpha \frac{\gamma_\alpha}{2}\left[2 L_\alpha\rho L_\alpha^\dagger - L_\alpha^\dagger L_\alpha \rho - \rho L_\alpha^\dagger L_\alpha\right].
\end{equation}
We now consider the unitarily transformed master equation
\begin{equation}\label{eq_LU_transformed}
  U^\dagger\left(\mathcal{L}_U\rho\right)U=\sum_\alpha \frac{\gamma_\alpha}{2}\left[2 \tilde{L}_\alpha\tilde{\rho} \tilde{L}_\alpha^\dagger - \tilde{L}_\alpha^\dagger \tilde{L}_\alpha \tilde{\rho} - \tilde{\rho} \tilde{L}_\alpha^\dagger \tilde{L}_\alpha\right],
\end{equation}
where we have defined $\tilde{\rho}\coloneq U^\dagger\rho U$ and $\tilde{L}_\alpha\coloneq U^\dagger L_\alpha U$. The right-hand side of Eq.~\eqref{eq_LU_transformed} is thus again a Lindblad superoperator, $U^\dagger(\mathcal{L}_U\rho)U\eqcolon \tilde{\mathcal{L}}\tilde{\rho}$. Now, since $\rho_\mathrm{ss}=U\rho_\mathrm{th}U^\dagger$ is the steady-state solution of $\mathcal{L}_U$, the state $\tilde{\rho}_\mathrm{ss}\coloneq U^\dagger\rho_\mathrm{ss}U= \rho_\mathrm{th}$ must be the steady state of $\tilde{\mathcal{L}}$. Hence, $\tilde{\mathcal{L}}$ has to be a thermal generator, i.e., $\tilde{\mathcal{L}}=\mathcal{L}_\mathrm{th}$, and therefore
\begin{equation}\label{eq_LU_Lth}
  U^\dagger(\mathcal{L}_U\rho)U=\mathcal{L}_\mathrm{th}\left(U^\dagger\rho U\right).
\end{equation}
Hence, the solution of $\dot\rho=\mathcal{L}_U\rho$ may be written as
\begin{equation}
  \rho(t)=U\left[e^{t\mathcal{L}_\mathrm{th}}\left(U^\dagger\rho_0 U\right)\right]U^\dagger.
\end{equation}

\par

If $H(t)$ is slowly varying in time and commutes with itself at all times, we have time-dependent $\gamma_\alpha(t)$ in Eq.~\eqref{eq_LU_Lalpha}~\cite{alicki1979quantum}. Since the above derivation does not depend on these rates, we have
\begin{equation}\label{eq_liouvillian_U_th_t}
  U^\dagger\left(\mathcal{L}_U(t)\rho(t)\right)U=\mathcal{L}_\mathrm{th}(t)\left(U^\dagger\rho(t) U\right).
\end{equation}
  
\section{Entropy change for time-dependent Hamiltonians}\label{app_time-dependent_Hamiltonian}

Equation~\eqref{eq_DeltaS_Qth} for a thermal bath was derived based on the alternative (dashed) path
\begin{equation}\label{eq_app_Lth_t_alternative_paths}
  \begin{tikzcd}[row sep=huge, column sep = 6em]
    \rho_0 \arrow[r,"\dot\rho(t)=\mathcal{L}_\mathrm{th}(t)\rho(t)","\mathcal{E}_\mathrm{d}"',mapsto] \arrow[d,"\mathrm{unitary}"',mapsto,bend right,dashed] & \rho_\mathrm{th}(\infty) \\ \varrho_0=\pi_0 \arrow[ru,"\dot\varrho(t)=\mathcal{L}_\mathrm{th}(t)\varrho(t)"',"\mathcal{E}_\mathrm{d}^\prime",mapsto,bend right,dashed] 
  \end{tikzcd}.
\end{equation}
The energies $\mathcal{E}_\mathrm{d}$ (along the original path) and $\mathcal{E}_\mathrm{d}^\prime$ (along the alternative path) are those that appear on the r.h.s.\ of the entropic inequalities~\eqref{eq_DeltaS_QdT_DeltaW} and~\eqref{eq_DeltaS_Qth}.

\par

The $\Sigma$-inequality for the situation where the invariant state is non-passive is given in Eq.~\eqref{eq_sigma_t_LU_integrated} and may be recast in the form
\begin{equation}\label{eq_eq_sigma_t_LU_integrated_thermal}
  \Delta\mathcal{S}\geq\frac{1}{T}\int_{0}^{\infty}\Tr\Big[U^\dagger[\mathcal{L}_U(t)\rho(t)]UH(t)\Big]\dd t.
\end{equation}
Owing to Eq.~\eqref{eq_liouvillian_U_th_t}, this inequality is equivalent to
\begin{equation}\label{eq_LU_Lth_integral}
  \Delta\mathcal{S}\geq\frac{1}{T}\int_{0}^{\infty}\Tr\Big[[\mathcal{L}_\mathrm{th}(t)\tilde\rho(t)]H(t)\Big]\dd t,
\end{equation}
where $\tilde\rho(t)\coloneq U^\dagger\rho(t)U$ and $\mathcal{L}_\mathrm{th}(t)$ is a thermal Liouvillian with the same temperature and the same $H(t)$ as in $\mathcal{L}_U(t)$. The problem of a state $\rho(t)$ that evolves subject to a non-thermal bath has thus been reduced to the problem of a state $\tilde\rho(t)$ that evolves according to a thermal bath. This is the situation considered in the original (solid) path in Eq.~\eqref{eq_app_Lth_t_alternative_paths} upon replacing $\rho(t)$ by $\tilde\rho(t)$ there. This yields again Eq.~\eqref{eq_DeltaS_Qth}, thus extending it to the case of a non-passive invariant state.

\par

In the general case that $\pi_\mathrm{ss}(t)$ is not a thermal state, inequality~\eqref{eq_eq_sigma_t_LU_integrated_thermal} is replaced by
\begin{equation}
  \Delta\mathcal{S}\geq-\kB\int_{0}^{\infty}\Tr\Big[U^\dagger[\mathcal{L}_U(t)\rho(t)]U\ln\pi_\mathrm{ss}(t)\Big]\dd t.
\end{equation}
One can then proceed as above, but $\mathcal{L}_\mathrm{th}(t)$ is then replaced by a ``passive'' Liouvillian $\mathcal{L}_\mathrm{pas}(t)$ whose invariant state is $\pi_\mathrm{ss}(t)$. The resulting inequality for $\Delta\mathcal{S}$ [the generalisation of Eq.~\eqref{eq_DeltaS_Qth}, i.e., the counterpart of Eq.~\eqref{eq_srel}] then reads,
\begin{equation}\label{eq_app_DeltaS_integral_passive}
  \Delta\mathcal{S}\geq -\kB\int_{0}^{\infty}\Tr\Big[[\mathcal{L}_\mathrm{pas}(t)\varrho(t)]\ln\pi_\mathrm{ss}(t)\Big]\dd t,
\end{equation}
where $\varrho(0)=\pi_0$. Note that the latter integral cannot be identified with energy transfer. Equation~\eqref{eq_app_DeltaS_integral_passive} holds also for the case of a passive invariant state $\rho_\mathrm{ss}(t)=\pi_\mathrm{ss}(t)$, where now $\mathcal{L}_\mathrm{pas}(t)=\mathcal{L}(t)$.

\section{Derivation of the efficiency bound}\label{app_efficiency}

Energy conservation [Eq.~\eqref{eq_first_law}] over a cycle yields
\begin{equation}
  \mathcal{E}_\mathrm{d,c}+\mathcal{E}_\mathrm{d,h}+W=0,
\end{equation}
where $\mathcal{E}_\mathrm{d,c}$ ($\mathcal{E}_\mathrm{d,h}$) is the dissipative energy change of the WM due to its interaction with the cold thermal (hot non-thermal) bath (Fig.~\ref{fig_engine}). As mentioned in the main text, we assume that the WM is thermal and hence passive prior to its interaction with the cold thermal bath.

\par

The efficiency of the engine is defined as the ratio of the extracted work to the invested energy (passive thermal energy and ergotropy) $\mathcal{E}_\mathrm{d,h}=\int_0^\infty\Tr[(\mathcal{L}_U(t)\rho(t))H(t)]\dd t$ provided by the non-thermal bath, yielding
\begin{equation}\label{eq_app_eta}
  \eta\coloneq\frac{-W}{\mathcal{E}_\mathrm{d,h}}=1+\frac{\mathcal{E}_\mathrm{d,c}}{\mathcal{E}_\mathrm{d,h}}.
\end{equation}
This expression holds for $\mathcal{E}_\mathrm{d,c}\leq0$ and $\mathcal{E}_\mathrm{d,h}\geq0$; see below a discussion of the opposite case. From condition~\eqref{eq_condition_gen} it then follows that
  \begin{equation}
    \mathcal{E}_\mathrm{d,c}\leq -\frac{T_\mathrm{c}}{T_\mathrm{h}}\mathcal{E}_\mathrm{d,h}^\prime.
  \end{equation}
Inserting this relation into~\eqref{eq_app_eta} yields the efficiency bound~\eqref{eq_etamax_gen}.

\par

The efficiency bound~\eqref{eq_etamax_gen} may be generalised to the case where the passive state of the working medium is not thermal after the interaction with the non-thermal bath. Condition~\eqref{eq_condition_gen} is then, following Eq.~\eqref{eq_app_DeltaS_integral_passive}, replaced by
  \begin{equation}
    \frac{\mathcal{E}_\mathrm{d,c}}{T_\mathrm{c}}-\kB\int_{0}^{\infty}\Tr\Big[[\mathcal{L}_\mathrm{pas}(t)\varrho(t)]\ln\pi_\mathrm{ss}(t)\Big]\dd t\leq 0
  \end{equation}
and we then find
\begin{equation}
  \eta\leq 1+\frac{\kB T_\mathrm{c}}{\mathcal{E}_\mathrm{d,h}}\int_{0}^{\infty}\Tr\Big[[\mathcal{L}_\mathrm{pas}(t)\varrho(t)]\ln\pi_\mathrm{ss}(t)\Big]\dd t,
\end{equation}
where the integral is evaluated for the energising stroke.

\par

If $\mathcal{E}_\mathrm{d,c}>0$ ($\mathcal{E}_\mathrm{d,h}^\prime<0$), then also the cold bath provides energy, which has to be taken into account in the efficiency. The latter now reads~\cite{niedenzu2016operation}
\begin{equation}\label{eq_app_eta_1}
  \eta=\frac{-W}{\mathcal{E}_\mathrm{d,h}+\mathcal{E}_\mathrm{d,c}}=\frac{\mathcal{E}_\mathrm{d,h}+\mathcal{E}_\mathrm{d,c}}{\mathcal{E}_\mathrm{d,h}+\mathcal{E}_\mathrm{d,c}}=1,
\end{equation}
which cannot be further restricted by any inequality for $\Delta\mathcal{S}$.

\par

We now derive the efficiency bound that follows from the reversibility condition~\eqref{eq_spohn_integrated}. The requirement of vanishing entropy change over a cycle then yields
\begin{equation}
  \frac{\mathcal{E}_\mathrm{d,c}}{T_\mathrm{c}}+\frac{\tilde{\mathcal{E}}_\mathrm{d,h}}{T_\mathrm{h}}\leq0,
\end{equation}
where $\tilde{\mathcal{E}}_\mathrm{d,h}$ [the integral in Eq.~\eqref{eq_sigma_t_LU_integrated}] is the energy change during the interaction with the thermal bath along the dashed path in Eq.~\eqref{eq_cd}. Consequently, according to this criterion the efficiency~\eqref{eq_app_eta} is bounded by
\begin{equation}\label{eq_eta_bound_spohn}
  \eta\leq 1-\frac{T_\mathrm{c}}{T_\mathrm{h}}\frac{\tilde{\mathcal{E}}_\mathrm{d,h}}{\mathcal{E}_\mathrm{d,h}}\eqcolon\eta_\Sigma.
\end{equation}
This bound surpasses $1$ if $\tilde{\mathcal{E}}_\mathrm{d,h}<0$, which, e.g., is the case if the bath is ``over-squeezed'': This means that, due to the excessive bath squeezing, the interaction with the thermal bath along the alternative path of Eq.~\eqref{eq_cd} decreases the energy while that with the non-thermal bath along the initial path increases it.

\par

If the Hamiltonian is constant during the energising stroke, then $\tilde{\mathcal{E}}_\mathrm{d,h}=\Delta E_\mathrm{pas,h}|_\mathrm{d}+\widetilde{\Delta\mathcal{W}|_\mathrm{d}}$, where $\widetilde{\Delta\mathcal{W}|_\mathrm{d}}\leq 0$ is the ergotropy lost to the effective thermal bath in the second step of the alternative path in Eq.~\eqref{eq_cd}. A comparison of Eq.~\eqref{eq_eta_bound_spohn} with our bound Eq.~\eqref{eq_etamax_otto} for a constant Hamiltonian then yields $\eta_\mathrm{max}^\mathrm{Otto}\leq\eta_\Sigma$.

\section{Maximal efficiency of multi-bath quantum engines}\label{app_multibath}

We consider a cycle operating between $N$ thermal baths (either heat sources or heat dumps) and $M$ non-thermal baths that are assumed to energise the engine. Namely, the non-thermal baths provide both passive energy and ergotropy to the working medium. As before (see main text and Appendix~\ref{app_efficiency}) we assume that the strokes are sufficiently long such that Eq.~\eqref{eq_DeltaS_Qth} is valid and that the ergotropy of the working medium is extracted before every stroke that involves a bath.

\par

For this situation, Eq.~\eqref{eq_condition_gen} can be generalised to
\begin{equation}\label{eq_app_condition_multibath}
  0\geq\sum_{i=1}^M \frac{\mathcal{E}_{\mathrm{d,h},i}^{\prime}}{T_{\mathrm{h},i}}+\sum_{\{1\leq i\leq N|\mathcal{E}_{\mathrm{d,}i}\geq 0\}}\frac{\mathcal{E}_{\mathrm{d,}i}}{T_i}+\sum_{\{1\leq i\leq N|\mathcal{E}_{\mathrm{d,}i}\leq 0\}}\frac{\mathcal{E}_{\mathrm{d},i}}{T_i}.
\end{equation}
Here the temperatures of the thermal baths are denoted by $T_i$ and the temperature parameters of the non-thermal baths by $T_{\mathrm{h},i}$. Note that under the assumptions made above $\mathcal{E}_{\mathrm{d,h},i}^{\prime}\geq0$ and that for thermal baths $\mathcal{E}_{\mathrm{d,}i}\equiv\mathcal{E}_{\mathrm{d,}i}^\prime$.

\par

By introducing the minimum and maximum temperatures $T_\mathrm{min}\leq\{T_i,T_{\mathrm{h},i}\}\leq T_\mathrm{max}$, we obtain~\cite{schwablbook}
\begin{multline}\label{eq_app_condition_multibath_minmaxtemp}
  0\geq\sum_{i=1}^M \frac{\mathcal{E}_{\mathrm{d,h},i}^{\prime}}{T_{\mathrm{h},i}}+\sum_{\{1\leq i\leq N|\mathcal{E}_{\mathrm{d,}i}^\prime\geq 0\}}\frac{\mathcal{E}_{\mathrm{d,}i}^\prime}{T_i}+\sum_{\{1\leq i\leq N|\mathcal{E}_{\mathrm{d,}i}\leq 0\}}\frac{\mathcal{E}_{\mathrm{d},i}}{T_i}\\\geq\frac{\sum_{i=1}^M \mathcal{E}_{\mathrm{d,h},i}^{\prime}+\sum_{\{i|\mathcal{E}_{\mathrm{d},i}^\prime\geq 0\}}\mathcal{E}_{\mathrm{d,}i}^\prime}{T_\mathrm{max}}+\frac{\sum_{\{i|\mathcal{E}_{\mathrm{d,}i}\leq 0\}}\mathcal{E}_{\mathrm{d,}i}}{T_\mathrm{min}}\\\eqcolon \frac{\mathcal{E}_\mathrm{d,in}^\prime}{T_\mathrm{max}}+\frac{\mathcal{E}_\mathrm{d,out}}{T_\mathrm{min}}.
\end{multline}
Hence, we have the relation
\begin{equation}\label{eq_app_Eout}
  \mathcal{E}_\mathrm{d,out}\leq-\frac{T_\mathrm{min}}{T_\mathrm{max}}\mathcal{E}^\prime_\mathrm{d,in}.
\end{equation}

\par

The efficiency of the multi-bath engine is
\begin{equation}\label{eq_app_eta_multibath}
  \eta=1+\frac{\mathcal{E}_\mathrm{d,out}}{\mathcal{E}_\mathrm{d,in}},
\end{equation}
where
\begin{equation}
  \mathcal{E}_\mathrm{d,in}\coloneq\sum_{i=1}^M \mathcal{E}_{\mathrm{d,h},i}+\sum_{\{1\leq i\leq N|\mathcal{E}_{\mathrm{d,}i}\geq 0\}}\mathcal{E}_{\mathrm{d,}i}
\end{equation}
is the total energy that the working medium obtained from the energising baths during a cycle. Owing to Eq.~\eqref{eq_app_Eout}, the efficiency~\eqref{eq_app_eta_multibath} is bounded by
\begin{equation}\label{eq_app_etamax_multibath}
  \eta\leq 1-\frac{T_\mathrm{min}}{T_\mathrm{max}}\frac{\mathcal{E}^\prime_\mathrm{d,in}}{\mathcal{E}_\mathrm{d,in}}.
\end{equation}
Note that the equality sign in Eq.~\eqref{eq_app_etamax_multibath} is only fulfilled if both equality signs in Eq.~\eqref{eq_app_condition_multibath_minmaxtemp} hold. In particular, Eq.~\eqref{eq_app_etamax_multibath} is a strict inequality in the multi-bath case, i.e., if more than two temperatures appear in Eq.~\eqref{eq_app_condition_multibath}.

\par

Inequality~\eqref{eq_app_etamax_multibath} is the generalisation of Eq.~\eqref{eq_etamax_gen} to more than one energising bath. The efficiency of multi-bath engines is thus always lower than the maximum efficiency of a two-bath engine that operates between a cold thermal bath at temperature $T_\mathrm{min}$ and a hot non-thermal bath at temperature parameter $T_\mathrm{max}$ which results in the same ratio $\mathcal{E}_\mathrm{d,in}^\prime$/$\mathcal{E}_\mathrm{d,in}$ of the input energies. This also holds in the case that the first equality sign in Eq.~\eqref{eq_app_condition_multibath_minmaxtemp} is fulfilled, which in the case of thermal baths corresponds to the second law and hence the reversibility condition.

\par

The efficiency bound~\eqref{eq_app_etamax_multibath} thus contains as a special case the fact that the efficiency of multi-bath heat engines (i.e., the case where all the baths are thermal such that $\mathcal{E}^\prime_\mathrm{d,in}\equiv\mathcal{E}_\mathrm{d,in}$) is always lower than the Carnot efficiency determined by the minimium and the maximum temperatures of the cycle, even if the cycle is reversible~\cite{schwablbook}. In this sense, our bound~\eqref{eq_etamax_gen} is universal.

\par

The above considerations hold for the case $\mathcal{E}_{\mathrm{d,h},i}^\prime\geq0$. As discussed in Appendix~\ref{app_efficiency} for the two-bath situation, in the case that $\mathcal{E}_\mathrm{d,h}^\prime<0$ the two-bath engine operates at efficiency $\eta=1$ [Eq.~\eqref{eq_app_eta_1}], which obviously cannot be surpassed by any engine powered by multiple thermal or non-thermal baths.

\section{Expressions used in Figure~\ref{fig_efficiency_otto}}\label{app_figures}

In Fig.~\ref{fig_efficiency_otto} we have used the energies
\begin{subequations}\label{eq_energies_figure}
  \begin{align}
    \mathcal{E}_\mathrm{d,h}&=\hbar\omega_\mathrm{h} (\nbar_\mathrm{h}+\Delta\nbar_\mathrm{h}-\nbar_\mathrm{c})\label{eq_energies_figure_a}\\
    \Delta E_\mathrm{pas,h}&=\hbar\omega_\mathrm{h} (\nbar_\mathrm{h}-\nbar_\mathrm{c})\label{eq_energies_figure_b}\\
    \tilde{\mathcal{E}}_\mathrm{d}&=\Delta E_\mathrm{pas,h}|_\mathrm{d}-\hbar\omega_\mathrm{h}\Delta\nbar_\mathrm{c}.\label{eq_energies_figure_c}
  \end{align}
\end{subequations}
Here $\omega_\mathrm{c}$ ($\omega_\mathrm{h}$) is the oscillator frequency before (after) the compression stroke. Furthermore, we have defined $\nbar_i=[\exp(\hbar\omega_i/[\kB T_i])-1]^{-1}$ and $\Delta\nbar_i=(2\nbar_i+1)\sinh^2(r)$ for $i\in\{\mathrm{c},\mathrm{h}\}$, where $r$ denotes the squeezing parameter~\cite{breuerbook}. Using the energies~\eqref{eq_energies_figure}, the efficiency bounds $\eta_\Sigma$ [Eq.~\eqref{eq_eta_bound_spohn}] and $\eta_\mathrm{max}$ [Eq.~\eqref{eq_etamax_otto}] then evaluate to
\begin{equation}
  \eta_\Sigma=1-\frac{T_\mathrm{c}}{T_\mathrm{h}}\frac{\nbar_\mathrm{h}-\nbar_\mathrm{c}-\Delta\nbar_\mathrm{c}}{\nbar_\mathrm{h}+\Delta\nbar_\mathrm{h}-\nbar_\mathrm{c}}
\end{equation}
and
\begin{equation}
  \eta_\mathrm{max}=1-\frac{T_\mathrm{c}}{T_\mathrm{h}}\frac{\nbar_\mathrm{h}-\nbar_\mathrm{c}}{\nbar_\mathrm{h}+\Delta\nbar_\mathrm{h}-\nbar_\mathrm{c}},
\end{equation}
respectively. Additionally, we have used the actual efficiency~\cite{niedenzu2016operation}
\begin{equation}
  \eta=1-\frac{(\nbar_\mathrm{h}-\nbar_\mathrm{c})\omega_\mathrm{c}}{(\nbar_\mathrm{h}+\Delta\nbar_\mathrm{h}-\nbar_\mathrm{c})\omega_\mathrm{h}},
\end{equation}
which is valid for $\mathcal{E}_\mathrm{d,c}\leq0$, i.e., $\nbar_\mathrm{c}\leq\nbar_\mathrm{h}$. For $\nbar_\mathrm{h}\leq\nbar_\mathrm{c}\leq\nbar_\mathrm{h}+\Delta\nbar_\mathrm{h}$ the efficiency evaluates to $\eta=1$. The machine acts as an engine for $\mathcal{E}_\mathrm{d,h}\geq0$, i.e., for $\nbar_\mathrm{h}+\Delta\nbar_\mathrm{h}\geq\nbar_\mathrm{c}$, which for the parameters of Fig.~\ref{fig_efficiency_otto} corresponds to $\omega_\mathrm{c}/\omega_\mathrm{h}\gtrsim 0.22$.


\begin{thebibliography}{78}%
\makeatletter
\providecommand \@ifxundefined [1]{%
 \@ifx{#1\undefined}
}%
\providecommand \@ifnum [1]{%
 \ifnum #1\expandafter \@firstoftwo
 \else \expandafter \@secondoftwo
 \fi
}%
\providecommand \@ifx [1]{%
 \ifx #1\expandafter \@firstoftwo
 \else \expandafter \@secondoftwo
 \fi
}%
\providecommand \natexlab [1]{#1}%
\providecommand \enquote  [1]{#1}%
\providecommand \bibnamefont  [1]{#1}%
\providecommand \bibfnamefont [1]{#1}%
\providecommand \citenamefont [1]{#1}%
\providecommand \href@noop [0]{\@secondoftwo}%
\providecommand \href [0]{\begingroup \@sanitize@url \@href}%
\providecommand \@href[1]{\@@startlink{#1}\@@href}%
\providecommand \@@href[1]{\endgroup#1\@@endlink}%
\providecommand \@sanitize@url [0]{\catcode `\\12\catcode `\$12\catcode
  `\&12\catcode `\#12\catcode `\^12\catcode `\_12\catcode `\%12\relax}%
\providecommand \@@startlink[1]{}%
\providecommand \@@endlink[0]{}%
\providecommand \url  [0]{\begingroup\@sanitize@url \@url }%
\providecommand \@url [1]{\endgroup\@href {#1}{\urlprefix }}%
\providecommand \urlprefix  [0]{URL }%
\providecommand \Eprint [0]{\href }%
\providecommand \doibase [0]{http://dx.doi.org/}%
\providecommand \selectlanguage [0]{\@gobble}%
\providecommand \bibinfo  [0]{\@secondoftwo}%
\providecommand \bibfield  [0]{\@secondoftwo}%
\providecommand \translation [1]{[#1]}%
\providecommand \BibitemOpen [0]{}%
\providecommand \bibitemStop [0]{}%
\providecommand \bibitemNoStop [0]{.\EOS\space}%
\providecommand \EOS [0]{\spacefactor3000\relax}%
\providecommand \BibitemShut  [1]{\csname bibitem#1\endcsname}%
\let\auto@bib@innerbib\@empty
\bibitem [{\citenamefont {Carnot}(1824)}]{carnotbook}%
  \BibitemOpen
  \bibfield  {author} {\bibinfo {author} {\bibfnamefont {S.}~\bibnamefont
  {Carnot}},\ }\href@noop {} {\emph {\bibinfo {title} {R\'eflexions sur la
  puissance motrice du feu et sur les machines propres \`a d\'evelopper cette
  puissance}}}\ (\bibinfo  {publisher} {Bachelier},\ \bibinfo {address}
  {Paris},\ \bibinfo {year} {1824})\BibitemShut {NoStop}%
\bibitem [{\citenamefont {Schwabl}(2006)}]{schwablbook}%
  \BibitemOpen
  \bibfield  {author} {\bibinfo {author} {\bibfnamefont {F.}~\bibnamefont
  {Schwabl}},\ }\href@noop {} {\emph {\bibinfo {title} {Statistical
  Mechanics}}},\ \bibinfo {edition} {2nd}\ ed.\ (\bibinfo  {publisher}
  {Springer-Verlag},\ \bibinfo {address} {Berlin Heidelberg},\ \bibinfo {year}
  {2006})\BibitemShut {NoStop}%
\bibitem [{\citenamefont {Kondepudi}\ and\ \citenamefont
  {Prigogine}(2015)}]{kondepudibook}%
  \BibitemOpen
  \bibfield  {author} {\bibinfo {author} {\bibfnamefont {D.}~\bibnamefont
  {Kondepudi}}\ and\ \bibinfo {author} {\bibfnamefont {I.}~\bibnamefont
  {Prigogine}},\ }\href@noop {} {\emph {\bibinfo {title} {Modern
  Thermodynamics}}},\ \bibinfo {edition} {2nd}\ ed.\ (\bibinfo  {publisher}
  {John Wiley \& Sons Ltd},\ \bibinfo {address} {Chichester},\ \bibinfo {year}
  {2015})\BibitemShut {NoStop}%
\bibitem [{\citenamefont {Clausius}(1865)}]{clausius1865verschiedene}%
  \BibitemOpen
  \bibfield  {author} {\bibinfo {author} {\bibfnamefont {R.}~\bibnamefont
  {Clausius}},\ }\enquote {\bibinfo {title} {Ueber verschiedene f\"ur die
  Anwendung bequeme Formen der Hauptgleichungen der mechanischen
  W\"armetheorie},}\ \href {\doibase 10.1002/andp.18652010702} {\bibfield
  {journal} {\bibinfo  {journal} {Ann. Phys.}\ }\textbf {\bibinfo {volume}
  {201}},\ \bibinfo {pages} {353} (\bibinfo {year} {1865})}\BibitemShut
  {NoStop}%
\bibitem [{\citenamefont {Callen}(1985)}]{callenbook}%
  \BibitemOpen
  \bibfield  {author} {\bibinfo {author} {\bibfnamefont {H.~B.}\ \bibnamefont
  {Callen}},\ }\href@noop {} {\emph {\bibinfo {title} {{Thermodynamics and an
  Introduction to Thermostatistics}}}},\ \bibinfo {edition} {2nd}\ ed.\
  (\bibinfo  {publisher} {John Wiley \& Sons, Inc.},\ \bibinfo {address} {New
  York},\ \bibinfo {year} {1985})\BibitemShut {NoStop}%
\bibitem [{\citenamefont {Scovil}\ and\ \citenamefont
  {Schulz-DuBois}(1959)}]{scovil1959three}%
  \BibitemOpen
  \bibfield  {author} {\bibinfo {author} {\bibfnamefont {H.~E.~D.}\
  \bibnamefont {Scovil}}\ and\ \bibinfo {author} {\bibfnamefont {E.~O.}\
  \bibnamefont {Schulz-DuBois}},\ }\enquote {\bibinfo {title} {Three-Level
  Masers as Heat Engines},}\ \href {\doibase 10.1103/PhysRevLett.2.262}
  {\bibfield  {journal} {\bibinfo  {journal} {Phys. Rev. Lett.}\ }\textbf
  {\bibinfo {volume} {2}},\ \bibinfo {pages} {262} (\bibinfo {year}
  {1959})}\BibitemShut {NoStop}%
\bibitem [{\citenamefont {Pusz}\ and\ \citenamefont
  {Woronowicz}(1978)}]{pusz1978passive}%
  \BibitemOpen
  \bibfield  {author} {\bibinfo {author} {\bibfnamefont {W.}~\bibnamefont
  {Pusz}}\ and\ \bibinfo {author} {\bibfnamefont {S.~L.}\ \bibnamefont
  {Woronowicz}},\ }\enquote {\bibinfo {title} {Passive states and KMS states
  for general quantum systems},}\ \href {\doibase 10.1007/BF01614224}
  {\bibfield  {journal} {\bibinfo  {journal} {Commun. Math. Phys.}\ }\textbf
  {\bibinfo {volume} {58}},\ \bibinfo {pages} {273} (\bibinfo {year}
  {1978})}\BibitemShut {NoStop}%
\bibitem [{\citenamefont {Lenard}(1978)}]{lenard1978thermodynamical}%
  \BibitemOpen
  \bibfield  {author} {\bibinfo {author} {\bibfnamefont {A.}~\bibnamefont
  {Lenard}},\ }\enquote {\bibinfo {title} {Thermodynamical proof of the Gibbs
  formula for elementary quantum systems},}\ \href {\doibase
  10.1007/BF01011769} {\bibfield  {journal} {\bibinfo  {journal} {J. Stat.
  Phys.}\ }\textbf {\bibinfo {volume} {19}},\ \bibinfo {pages} {575} (\bibinfo
  {year} {1978})}\BibitemShut {NoStop}%
\bibitem [{\citenamefont {Alicki}(1979)}]{alicki1979quantum}%
  \BibitemOpen
  \bibfield  {author} {\bibinfo {author} {\bibfnamefont {R.}~\bibnamefont
  {Alicki}},\ }\enquote {\bibinfo {title} {The quantum open system as a model
  of the heat engine},}\ \href {\doibase 10.1088/0305-4470/12/5/007} {\bibfield
   {journal} {\bibinfo  {journal} {J. Phys. A}\ }\textbf {\bibinfo {volume}
  {12}},\ \bibinfo {pages} {L103} (\bibinfo {year} {1979})}\BibitemShut
  {NoStop}%
\bibitem [{\citenamefont {Scully}\ \emph {et~al.}(2003)\citenamefont {Scully},
  \citenamefont {Zubairy}, \citenamefont {Agarwal},\ and\ \citenamefont
  {Walther}}]{scully2003extracting}%
  \BibitemOpen
  \bibfield  {author} {\bibinfo {author} {\bibfnamefont {M.~O.}\ \bibnamefont
  {Scully}}, \bibinfo {author} {\bibfnamefont {M.~S.}\ \bibnamefont {Zubairy}},
  \bibinfo {author} {\bibfnamefont {G.~S.}\ \bibnamefont {Agarwal}}, \ and\
  \bibinfo {author} {\bibfnamefont {H.}~\bibnamefont {Walther}},\ }\enquote
  {\bibinfo {title} {Extracting Work from a Single Heat Bath via Vanishing
  Quantum Coherence},}\ \href {\doibase 10.1126/science.1078955} {\bibfield
  {journal} {\bibinfo  {journal} {Science}\ }\textbf {\bibinfo {volume}
  {299}},\ \bibinfo {pages} {862} (\bibinfo {year} {2003})}\BibitemShut
  {NoStop}%
\bibitem [{\citenamefont {Allahverdyan}\ \emph {et~al.}(2004)\citenamefont
  {Allahverdyan}, \citenamefont {Balian},\ and\ \citenamefont
  {Nieuwenhuizen}}]{allahverdyan2004maximal}%
  \BibitemOpen
  \bibfield  {author} {\bibinfo {author} {\bibfnamefont {A.~E.}\ \bibnamefont
  {Allahverdyan}}, \bibinfo {author} {\bibfnamefont {R.}~\bibnamefont
  {Balian}}, \ and\ \bibinfo {author} {\bibfnamefont {T.~M.}\ \bibnamefont
  {Nieuwenhuizen}},\ }\enquote {\bibinfo {title} {Maximal work extraction from
  finite quantum systems},}\ \href {\doibase 10.1209/epl/i2004-10101-2}
  {\bibfield  {journal} {\bibinfo  {journal} {EPL (Europhys. Lett.)}\ }\textbf
  {\bibinfo {volume} {67}},\ \bibinfo {pages} {565} (\bibinfo {year}
  {2004})}\BibitemShut {NoStop}%
\bibitem [{\citenamefont {Erez}\ \emph {et~al.}(2008)\citenamefont {Erez},
  \citenamefont {Gordon}, \citenamefont {Nest},\ and\ \citenamefont
  {Kurizki}}]{erez2008thermodynamic}%
  \BibitemOpen
  \bibfield  {author} {\bibinfo {author} {\bibfnamefont {N.}~\bibnamefont
  {Erez}}, \bibinfo {author} {\bibfnamefont {G.}~\bibnamefont {Gordon}},
  \bibinfo {author} {\bibfnamefont {M.}~\bibnamefont {Nest}}, \ and\ \bibinfo
  {author} {\bibfnamefont {G.}~\bibnamefont {Kurizki}},\ }\enquote {\bibinfo
  {title} {Thermodynamic control by frequent quantum measurements},}\ \href
  {\doibase 10.1038/nature06873} {\bibfield  {journal} {\bibinfo  {journal}
  {Nature}\ }\textbf {\bibinfo {volume} {452}},\ \bibinfo {pages} {724}
  (\bibinfo {year} {2008})}\BibitemShut {NoStop}%
\bibitem [{\citenamefont {Del~Rio}\ \emph {et~al.}(2011)\citenamefont
  {Del~Rio}, \citenamefont {{\AA}berg}, \citenamefont {Renner}, \citenamefont
  {Dahlsten},\ and\ \citenamefont {Vedral}}]{delrio2011thermodynamic}%
  \BibitemOpen
  \bibfield  {author} {\bibinfo {author} {\bibfnamefont {L.}~\bibnamefont
  {Del~Rio}}, \bibinfo {author} {\bibfnamefont {J.}~\bibnamefont {{\AA}berg}},
  \bibinfo {author} {\bibfnamefont {R.}~\bibnamefont {Renner}}, \bibinfo
  {author} {\bibfnamefont {O.}~\bibnamefont {Dahlsten}}, \ and\ \bibinfo
  {author} {\bibfnamefont {V.}~\bibnamefont {Vedral}},\ }\enquote {\bibinfo
  {title} {The thermodynamic meaning of negative entropy},}\ \href {\doibase
  10.1038/nature10123} {\bibfield  {journal} {\bibinfo  {journal} {Nature}\
  }\textbf {\bibinfo {volume} {474}},\ \bibinfo {pages} {61} (\bibinfo {year}
  {2011})}\BibitemShut {NoStop}%
\bibitem [{\citenamefont {Horodecki}\ and\ \citenamefont
  {Oppenheim}(2013)}]{horodecki2013fundamental}%
  \BibitemOpen
  \bibfield  {author} {\bibinfo {author} {\bibfnamefont {M.}~\bibnamefont
  {Horodecki}}\ and\ \bibinfo {author} {\bibfnamefont {J.}~\bibnamefont
  {Oppenheim}},\ }\enquote {\bibinfo {title} {Fundamental limitations for
  quantum and nanoscale thermodynamics},}\ \href {\doibase 10.1038/ncomms3059}
  {\bibfield  {journal} {\bibinfo  {journal} {Nat. Commun.}\ }\textbf {\bibinfo
  {volume} {4}},\ \bibinfo {pages} {2059} (\bibinfo {year} {2013})}\BibitemShut
  {NoStop}%
\bibitem [{\citenamefont {Correa}\ \emph {et~al.}(2014)\citenamefont {Correa},
  \citenamefont {Palao}, \citenamefont {Alonso},\ and\ \citenamefont
  {Adesso}}]{correa2014quantum}%
  \BibitemOpen
  \bibfield  {author} {\bibinfo {author} {\bibfnamefont {L.~A.}\ \bibnamefont
  {Correa}}, \bibinfo {author} {\bibfnamefont {J.~P.}\ \bibnamefont {Palao}},
  \bibinfo {author} {\bibfnamefont {D.}~\bibnamefont {Alonso}}, \ and\ \bibinfo
  {author} {\bibfnamefont {G.}~\bibnamefont {Adesso}},\ }\enquote {\bibinfo
  {title} {Quantum-enhanced absorption refrigerators},}\ \href {\doibase
  10.1038/srep03949} {\bibfield  {journal} {\bibinfo  {journal} {Sci. Rep.}\
  }\textbf {\bibinfo {volume} {4}},\ \bibinfo {pages} {3949} (\bibinfo {year}
  {2014})}\BibitemShut {NoStop}%
\bibitem [{\citenamefont {Skrzypczyk}\ \emph {et~al.}(2014)\citenamefont
  {Skrzypczyk}, \citenamefont {Short},\ and\ \citenamefont
  {Popescu}}]{skrzypczyk2014work}%
  \BibitemOpen
  \bibfield  {author} {\bibinfo {author} {\bibfnamefont {P.}~\bibnamefont
  {Skrzypczyk}}, \bibinfo {author} {\bibfnamefont {A.~J.}\ \bibnamefont
  {Short}}, \ and\ \bibinfo {author} {\bibfnamefont {S.}~\bibnamefont
  {Popescu}},\ }\enquote {\bibinfo {title} {Work extraction and thermodynamics
  for individual quantum systems},}\ \href {\doibase 10.1038/ncomms5185}
  {\bibfield  {journal} {\bibinfo  {journal} {Nat. Commun.}\ }\textbf {\bibinfo
  {volume} {5}},\ \bibinfo {pages} {4185} (\bibinfo {year} {2014})}\BibitemShut
  {NoStop}%
\bibitem [{\citenamefont {Brand{\~a}o}\ \emph {et~al.}(2015)\citenamefont
  {Brand{\~a}o}, \citenamefont {Horodecki}, \citenamefont {Ng}, \citenamefont
  {Oppenheim},\ and\ \citenamefont {Wehner}}]{brandao2015second}%
  \BibitemOpen
  \bibfield  {author} {\bibinfo {author} {\bibfnamefont {F.}~\bibnamefont
  {Brand{\~a}o}}, \bibinfo {author} {\bibfnamefont {M.}~\bibnamefont
  {Horodecki}}, \bibinfo {author} {\bibfnamefont {N.}~\bibnamefont {Ng}},
  \bibinfo {author} {\bibfnamefont {J.}~\bibnamefont {Oppenheim}}, \ and\
  \bibinfo {author} {\bibfnamefont {S.}~\bibnamefont {Wehner}},\ }\enquote
  {\bibinfo {title} {The second laws of quantum thermodynamics},}\ \href
  {\doibase 10.1073/pnas.1411728112} {\bibfield  {journal} {\bibinfo  {journal}
  {Proc. Natl. Acad. Sci.}\ }\textbf {\bibinfo {volume} {112}},\ \bibinfo
  {pages} {3275} (\bibinfo {year} {2015})}\BibitemShut {NoStop}%
\bibitem [{\citenamefont {Pekola}(2015)}]{pekola2015towards}%
  \BibitemOpen
  \bibfield  {author} {\bibinfo {author} {\bibfnamefont {J.~P.}\ \bibnamefont
  {Pekola}},\ }\enquote {\bibinfo {title} {Towards quantum thermodynamics in
  electronic circuits},}\ \href {\doibase 10.1038/nphys3169} {\bibfield
  {journal} {\bibinfo  {journal} {Nat. Phys.}\ }\textbf {\bibinfo {volume}
  {11}},\ \bibinfo {pages} {118} (\bibinfo {year} {2015})}\BibitemShut
  {NoStop}%
\bibitem [{\citenamefont {Uzdin}\ \emph {et~al.}(2015)\citenamefont {Uzdin},
  \citenamefont {Levy},\ and\ \citenamefont {Kosloff}}]{uzdin2015equivalence}%
  \BibitemOpen
  \bibfield  {author} {\bibinfo {author} {\bibfnamefont {R.}~\bibnamefont
  {Uzdin}}, \bibinfo {author} {\bibfnamefont {A.}~\bibnamefont {Levy}}, \ and\
  \bibinfo {author} {\bibfnamefont {R.}~\bibnamefont {Kosloff}},\ }\enquote
  {\bibinfo {title} {Equivalence of Quantum Heat Machines, and
  Quantum-Thermodynamic Signatures},}\ \href {\doibase
  10.1103/PhysRevX.5.031044} {\bibfield  {journal} {\bibinfo  {journal} {Phys.
  Rev. X}\ }\textbf {\bibinfo {volume} {5}},\ \bibinfo {pages} {031044}
  (\bibinfo {year} {2015})}\BibitemShut {NoStop}%
\bibitem [{\citenamefont {Campisi}\ and\ \citenamefont
  {Fazio}(2016)}]{campisi2016power}%
  \BibitemOpen
  \bibfield  {author} {\bibinfo {author} {\bibfnamefont {M.}~\bibnamefont
  {Campisi}}\ and\ \bibinfo {author} {\bibfnamefont {R.}~\bibnamefont
  {Fazio}},\ }\enquote {\bibinfo {title} {The power of a critical heat
  engine},}\ \href {\doibase 10.1038/ncomms11895} {\bibfield  {journal}
  {\bibinfo  {journal} {Nat. Commun.}\ }\textbf {\bibinfo {volume} {7}},\
  \bibinfo {pages} {11895} (\bibinfo {year} {2016})}\BibitemShut {NoStop}%
\bibitem [{\citenamefont {Ro{\ss}nagel}\ \emph {et~al.}(2016)\citenamefont
  {Ro{\ss}nagel}, \citenamefont {Dawkins}, \citenamefont {Tolazzi},
  \citenamefont {Abah}, \citenamefont {Lutz}, \citenamefont {Schmidt-Kaler},\
  and\ \citenamefont {Singer}}]{rossnagel2016single}%
  \BibitemOpen
  \bibfield  {author} {\bibinfo {author} {\bibfnamefont {J.}~\bibnamefont
  {Ro{\ss}nagel}}, \bibinfo {author} {\bibfnamefont {S.~T.}\ \bibnamefont
  {Dawkins}}, \bibinfo {author} {\bibfnamefont {K.~N.}\ \bibnamefont
  {Tolazzi}}, \bibinfo {author} {\bibfnamefont {O.}~\bibnamefont {Abah}},
  \bibinfo {author} {\bibfnamefont {E.}~\bibnamefont {Lutz}}, \bibinfo {author}
  {\bibfnamefont {F.}~\bibnamefont {Schmidt-Kaler}}, \ and\ \bibinfo {author}
  {\bibfnamefont {K.}~\bibnamefont {Singer}},\ }\enquote {\bibinfo {title} {A
  single-atom heat engine},}\ \href {\doibase 10.1126/science.aad6320}
  {\bibfield  {journal} {\bibinfo  {journal} {Science}\ }\textbf {\bibinfo
  {volume} {352}},\ \bibinfo {pages} {325} (\bibinfo {year}
  {2016})}\BibitemShut {NoStop}%
\bibitem [{\citenamefont {Kosloff}(2013)}]{kosloff2013quantum}%
  \BibitemOpen
  \bibfield  {author} {\bibinfo {author} {\bibfnamefont {R.}~\bibnamefont
  {Kosloff}},\ }\enquote {\bibinfo {title} {Quantum Thermodynamics: A Dynamical
  Viewpoint},}\ \href {\doibase 10.3390/e15062100} {\bibfield  {journal}
  {\bibinfo  {journal} {Entropy}\ }\textbf {\bibinfo {volume} {15}},\ \bibinfo
  {pages} {2100} (\bibinfo {year} {2013})}\BibitemShut {NoStop}%
\bibitem [{\citenamefont {Gelbwaser-Klimovsky}\ \emph
  {et~al.}(2015)\citenamefont {Gelbwaser-Klimovsky}, \citenamefont {Niedenzu},\
  and\ \citenamefont {Kurizki}}]{gelbwaser2015thermodynamics}%
  \BibitemOpen
  \bibfield  {author} {\bibinfo {author} {\bibfnamefont {D.}~\bibnamefont
  {Gelbwaser-Klimovsky}}, \bibinfo {author} {\bibfnamefont {W.}~\bibnamefont
  {Niedenzu}}, \ and\ \bibinfo {author} {\bibfnamefont {G.}~\bibnamefont
  {Kurizki}},\ }\enquote {\bibinfo {title} {Thermodynamics of Quantum Systems
  Under Dynamical Control},}\ \href {\doibase 10.1016/bs.aamop.2015.07.002}
  {\bibfield  {journal} {\bibinfo  {journal} {Adv. At. Mol. Opt. Phys.}\
  }\textbf {\bibinfo {volume} {64}},\ \bibinfo {pages} {329} (\bibinfo {year}
  {2015})}\BibitemShut {NoStop}%
\bibitem [{\citenamefont {Goold}\ \emph {et~al.}(2016)\citenamefont {Goold},
  \citenamefont {Huber}, \citenamefont {Riera}, \citenamefont {del Rio},\ and\
  \citenamefont {Skrzypczyk}}]{goold2016role}%
  \BibitemOpen
  \bibfield  {author} {\bibinfo {author} {\bibfnamefont {J.}~\bibnamefont
  {Goold}}, \bibinfo {author} {\bibfnamefont {M.}~\bibnamefont {Huber}},
  \bibinfo {author} {\bibfnamefont {A.}~\bibnamefont {Riera}}, \bibinfo
  {author} {\bibfnamefont {L.}~\bibnamefont {del Rio}}, \ and\ \bibinfo
  {author} {\bibfnamefont {P.}~\bibnamefont {Skrzypczyk}},\ }\enquote {\bibinfo
  {title} {The role of quantum information in thermodynamics---a topical
  review},}\ \href {\doibase 10.1088/1751-8113/49/14/143001} {\bibfield
  {journal} {\bibinfo  {journal} {J. Phys. A}\ }\textbf {\bibinfo {volume}
  {49}},\ \bibinfo {pages} {143001} (\bibinfo {year} {2016})}\BibitemShut
  {NoStop}%
\bibitem [{\citenamefont {Vinjanampathy}\ and\ \citenamefont
  {Anders}(2016)}]{vinjanampathy2016quantum}%
  \BibitemOpen
  \bibfield  {author} {\bibinfo {author} {\bibfnamefont {S.}~\bibnamefont
  {Vinjanampathy}}\ and\ \bibinfo {author} {\bibfnamefont {J.}~\bibnamefont
  {Anders}},\ }\enquote {\bibinfo {title} {Quantum thermodynamics},}\ \href
  {\doibase 10.1080/00107514.2016.1201896} {\bibfield  {journal} {\bibinfo
  {journal} {Contemp. Phys.}\ }\textbf {\bibinfo {volume} {57}},\ \bibinfo
  {pages} {1} (\bibinfo {year} {2016})}\BibitemShut {NoStop}%
\bibitem [{\citenamefont {Kosloff}\ and\ \citenamefont
  {Rezek}(2017)}]{kosloff2017quantum}%
  \BibitemOpen
  \bibfield  {author} {\bibinfo {author} {\bibfnamefont {R.}~\bibnamefont
  {Kosloff}}\ and\ \bibinfo {author} {\bibfnamefont {Y.}~\bibnamefont
  {Rezek}},\ }\enquote {\bibinfo {title} {The Quantum Harmonic Otto Cycle},}\
  \href {\doibase 10.3390/e19040136} {\bibfield  {journal} {\bibinfo  {journal}
  {Entropy}\ }\textbf {\bibinfo {volume} {19}},\ \bibinfo {pages} {136}
  (\bibinfo {year} {2017})}\BibitemShut {NoStop}%
\bibitem [{\citenamefont {Dillenschneider}\ and\ \citenamefont
  {Lutz}(2009)}]{dillenschneider2009energetics}%
  \BibitemOpen
  \bibfield  {author} {\bibinfo {author} {\bibfnamefont {R.}~\bibnamefont
  {Dillenschneider}}\ and\ \bibinfo {author} {\bibfnamefont {E.}~\bibnamefont
  {Lutz}},\ }\enquote {\bibinfo {title} {Energetics of quantum correlations},}\
  \href {\doibase 10.1209/0295-5075/88/50003} {\bibfield  {journal} {\bibinfo
  {journal} {EPL (Europhys. Lett.)}\ }\textbf {\bibinfo {volume} {88}},\
  \bibinfo {pages} {50003} (\bibinfo {year} {2009})}\BibitemShut {NoStop}%
\bibitem [{\citenamefont {Huang}\ \emph {et~al.}(2012)\citenamefont {Huang},
  \citenamefont {Wang},\ and\ \citenamefont {Yi}}]{huang2012effects}%
  \BibitemOpen
  \bibfield  {author} {\bibinfo {author} {\bibfnamefont {X.~L.}\ \bibnamefont
  {Huang}}, \bibinfo {author} {\bibfnamefont {T.}~\bibnamefont {Wang}}, \ and\
  \bibinfo {author} {\bibfnamefont {X.~X.}\ \bibnamefont {Yi}},\ }\enquote
  {\bibinfo {title} {Effects of reservoir squeezing on quantum systems and work
  extraction},}\ \href {\doibase 10.1103/PhysRevE.86.051105} {\bibfield
  {journal} {\bibinfo  {journal} {Phys. Rev. E}\ }\textbf {\bibinfo {volume}
  {86}},\ \bibinfo {pages} {051105} (\bibinfo {year} {2012})}\BibitemShut
  {NoStop}%
\bibitem [{\citenamefont {Abah}\ and\ \citenamefont
  {Lutz}(2014)}]{abah2014efficiency}%
  \BibitemOpen
  \bibfield  {author} {\bibinfo {author} {\bibfnamefont {O.}~\bibnamefont
  {Abah}}\ and\ \bibinfo {author} {\bibfnamefont {E.}~\bibnamefont {Lutz}},\
  }\enquote {\bibinfo {title} {Efficiency of heat engines coupled to
  nonequilibrium reservoirs},}\ \href {\doibase 10.1209/0295-5075/106/20001}
  {\bibfield  {journal} {\bibinfo  {journal} {EPL (Europhys. Lett.)}\ }\textbf
  {\bibinfo {volume} {106}},\ \bibinfo {pages} {20001} (\bibinfo {year}
  {2014})}\BibitemShut {NoStop}%
\bibitem [{\citenamefont {Ro\ss{}nagel}\ \emph {et~al.}(2014)\citenamefont
  {Ro\ss{}nagel}, \citenamefont {Abah}, \citenamefont {Schmidt-Kaler},
  \citenamefont {Singer},\ and\ \citenamefont {Lutz}}]{rossnagel2014nanoscale}%
  \BibitemOpen
  \bibfield  {author} {\bibinfo {author} {\bibfnamefont {J.}~\bibnamefont
  {Ro\ss{}nagel}}, \bibinfo {author} {\bibfnamefont {O.}~\bibnamefont {Abah}},
  \bibinfo {author} {\bibfnamefont {F.}~\bibnamefont {Schmidt-Kaler}}, \bibinfo
  {author} {\bibfnamefont {K.}~\bibnamefont {Singer}}, \ and\ \bibinfo {author}
  {\bibfnamefont {E.}~\bibnamefont {Lutz}},\ }\enquote {\bibinfo {title}
  {Nanoscale Heat Engine Beyond the Carnot Limit},}\ \href {\doibase
  10.1103/PhysRevLett.112.030602} {\bibfield  {journal} {\bibinfo  {journal}
  {Phys. Rev. Lett.}\ }\textbf {\bibinfo {volume} {112}},\ \bibinfo {pages}
  {030602} (\bibinfo {year} {2014})}\BibitemShut {NoStop}%
\bibitem [{\citenamefont {Hardal}\ and\ \citenamefont
  {M{\"u}stecapl{\i}o{\u{g}}lu}(2015)}]{hardal2015superradiant}%
  \BibitemOpen
  \bibfield  {author} {\bibinfo {author} {\bibfnamefont {A.~{\"U}.~C.}\
  \bibnamefont {Hardal}}\ and\ \bibinfo {author} {\bibfnamefont {{\"O}.~E.}\
  \bibnamefont {M{\"u}stecapl{\i}o{\u{g}}lu}},\ }\enquote {\bibinfo {title}
  {Superradiant Quantum Heat Engine},}\ \href {\doibase 10.1038/srep12953}
  {\bibfield  {journal} {\bibinfo  {journal} {Sci. Rep.}\ }\textbf {\bibinfo
  {volume} {5}},\ \bibinfo {pages} {12953} (\bibinfo {year}
  {2015})}\BibitemShut {NoStop}%
\bibitem [{\citenamefont {Niedenzu}\ \emph {et~al.}(2016)\citenamefont
  {Niedenzu}, \citenamefont {Gelbwaser-Klimovsky}, \citenamefont {Kofman},\
  and\ \citenamefont {Kurizki}}]{niedenzu2016operation}%
  \BibitemOpen
  \bibfield  {author} {\bibinfo {author} {\bibfnamefont {W.}~\bibnamefont
  {Niedenzu}}, \bibinfo {author} {\bibfnamefont {D.}~\bibnamefont
  {Gelbwaser-Klimovsky}}, \bibinfo {author} {\bibfnamefont {A.~G.}\
  \bibnamefont {Kofman}}, \ and\ \bibinfo {author} {\bibfnamefont
  {G.}~\bibnamefont {Kurizki}},\ }\enquote {\bibinfo {title} {On the operation
  of machines powered by quantum non-thermal baths},}\ \href {\doibase
  10.1088/1367-2630/18/8/083012} {\bibfield  {journal} {\bibinfo  {journal}
  {New J. Phys.}\ }\textbf {\bibinfo {volume} {18}},\ \bibinfo {pages} {083012}
  (\bibinfo {year} {2016})}\BibitemShut {NoStop}%
\bibitem [{\citenamefont {Manzano}\ \emph {et~al.}(2016)\citenamefont
  {Manzano}, \citenamefont {Galve}, \citenamefont {Zambrini},\ and\
  \citenamefont {Parrondo}}]{manzano2016entropy}%
  \BibitemOpen
  \bibfield  {author} {\bibinfo {author} {\bibfnamefont {G.}~\bibnamefont
  {Manzano}}, \bibinfo {author} {\bibfnamefont {F.}~\bibnamefont {Galve}},
  \bibinfo {author} {\bibfnamefont {R.}~\bibnamefont {Zambrini}}, \ and\
  \bibinfo {author} {\bibfnamefont {J.~M.~R.}\ \bibnamefont {Parrondo}},\
  }\enquote {\bibinfo {title} {Entropy production and thermodynamic power of
  the squeezed thermal reservoir},}\ \href {\doibase
  10.1103/PhysRevE.93.052120} {\bibfield  {journal} {\bibinfo  {journal} {Phys.
  Rev. E}\ }\textbf {\bibinfo {volume} {93}},\ \bibinfo {pages} {052120}
  (\bibinfo {year} {2016})}\BibitemShut {NoStop}%
\bibitem [{\citenamefont {Klaers}\ \emph {et~al.}(2017)\citenamefont {Klaers},
  \citenamefont {Faelt}, \citenamefont {Imamoglu},\ and\ \citenamefont
  {Togan}}]{klaers2017squeezed}%
  \BibitemOpen
  \bibfield  {author} {\bibinfo {author} {\bibfnamefont {J.}~\bibnamefont
  {Klaers}}, \bibinfo {author} {\bibfnamefont {S.}~\bibnamefont {Faelt}},
  \bibinfo {author} {\bibfnamefont {A.}~\bibnamefont {Imamoglu}}, \ and\
  \bibinfo {author} {\bibfnamefont {E.}~\bibnamefont {Togan}},\ }\enquote
  {\bibinfo {title} {Squeezed Thermal Reservoirs as a Resource for a
  Nanomechanical Engine beyond the Carnot Limit},}\ \href {\doibase
  10.1103/PhysRevX.7.031044} {\bibfield  {journal} {\bibinfo  {journal} {Phys.
  Rev. X}\ }\textbf {\bibinfo {volume} {7}},\ \bibinfo {pages} {031044}
  (\bibinfo {year} {2017})}\BibitemShut {NoStop}%
\bibitem [{\citenamefont {Agarwalla}\ \emph {et~al.}(2017)\citenamefont
  {Agarwalla}, \citenamefont {Jiang},\ and\ \citenamefont
  {Segal}}]{agarwalla2017quantum}%
  \BibitemOpen
  \bibfield  {author} {\bibinfo {author} {\bibfnamefont {B.~K.}\ \bibnamefont
  {Agarwalla}}, \bibinfo {author} {\bibfnamefont {J.-H.}\ \bibnamefont
  {Jiang}}, \ and\ \bibinfo {author} {\bibfnamefont {D.}~\bibnamefont
  {Segal}},\ }\enquote {\bibinfo {title} {Quantum efficiency bound for
  continuous heat engines coupled to noncanonical reservoirs},}\ \href
  {\doibase 10.1103/PhysRevB.96.104304} {\bibfield  {journal} {\bibinfo
  {journal} {Phys. Rev. B}\ }\textbf {\bibinfo {volume} {96}},\ \bibinfo
  {pages} {104304} (\bibinfo {year} {2017})}\BibitemShut {NoStop}%
\bibitem [{\citenamefont {Da{\u{g}}}\ \emph {et~al.}(2016)\citenamefont
  {Da{\u{g}}}, \citenamefont {Niedenzu}, \citenamefont
  {M{\"u}stecapl{\i}o{\u{g}}lu},\ and\ \citenamefont
  {Kurizki}}]{dag2016multiatom}%
  \BibitemOpen
  \bibfield  {author} {\bibinfo {author} {\bibfnamefont {C.~B.}\ \bibnamefont
  {Da{\u{g}}}}, \bibinfo {author} {\bibfnamefont {W.}~\bibnamefont {Niedenzu}},
  \bibinfo {author} {\bibfnamefont {{\"O}.~E.}\ \bibnamefont
  {M{\"u}stecapl{\i}o{\u{g}}lu}}, \ and\ \bibinfo {author} {\bibfnamefont
  {G.}~\bibnamefont {Kurizki}},\ }\enquote {\bibinfo {title} {Multiatom Quantum
  Coherences in Micromasers as Fuel for Thermal and Nonthermal Machines},}\
  \href {\doibase 10.3390/e18070244} {\bibfield  {journal} {\bibinfo  {journal}
  {Entropy}\ }\textbf {\bibinfo {volume} {18}},\ \bibinfo {pages} {244}
  (\bibinfo {year} {2016})}\BibitemShut {NoStop}%
\bibitem [{\citenamefont {Alicki}\ \emph {et~al.}(2004)\citenamefont {Alicki},
  \citenamefont {Horodecki}, \citenamefont {Horodecki},\ and\ \citenamefont
  {Horodecki}}]{alicki2004thermodynamics}%
  \BibitemOpen
  \bibfield  {author} {\bibinfo {author} {\bibfnamefont {R.}~\bibnamefont
  {Alicki}}, \bibinfo {author} {\bibfnamefont {M.}~\bibnamefont {Horodecki}},
  \bibinfo {author} {\bibfnamefont {P.}~\bibnamefont {Horodecki}}, \ and\
  \bibinfo {author} {\bibfnamefont {R.}~\bibnamefont {Horodecki}},\ }\enquote
  {\bibinfo {title} {Thermodynamics of Quantum Information Systems —
  Hamiltonian Description},}\ \href {\doibase
  10.1023/B:OPSY.0000047566.72717.71} {\bibfield  {journal} {\bibinfo
  {journal} {Open Syst. Inf. Dyn.}\ }\textbf {\bibinfo {volume} {11}},\
  \bibinfo {pages} {205} (\bibinfo {year} {2004})}\BibitemShut {NoStop}%
\bibitem [{\citenamefont {Boukobza}\ and\ \citenamefont
  {Tannor}(2007)}]{boukobza2007three}%
  \BibitemOpen
  \bibfield  {author} {\bibinfo {author} {\bibfnamefont {E.}~\bibnamefont
  {Boukobza}}\ and\ \bibinfo {author} {\bibfnamefont {D.~J.}\ \bibnamefont
  {Tannor}},\ }\enquote {\bibinfo {title} {Three-Level Systems as Amplifiers
  and Attenuators: A Thermodynamic Analysis},}\ \href {\doibase
  10.1103/PhysRevLett.98.240601} {\bibfield  {journal} {\bibinfo  {journal}
  {Phys. Rev. Lett.}\ }\textbf {\bibinfo {volume} {98}},\ \bibinfo {pages}
  {240601} (\bibinfo {year} {2007})}\BibitemShut {NoStop}%
\bibitem [{\citenamefont {Parrondo}\ \emph {et~al.}(2009)\citenamefont
  {Parrondo}, \citenamefont {den Broeck},\ and\ \citenamefont
  {Kawai}}]{parrondo2009entropy}%
  \BibitemOpen
  \bibfield  {author} {\bibinfo {author} {\bibfnamefont {J.~M.~R.}\
  \bibnamefont {Parrondo}}, \bibinfo {author} {\bibfnamefont {C.~V.}\
  \bibnamefont {den Broeck}}, \ and\ \bibinfo {author} {\bibfnamefont
  {R.}~\bibnamefont {Kawai}},\ }\enquote {\bibinfo {title} {Entropy production
  and the arrow of time},}\ \href {\doibase 10.1088/1367-2630/11/7/073008}
  {\bibfield  {journal} {\bibinfo  {journal} {New J. Phys.}\ }\textbf {\bibinfo
  {volume} {11}},\ \bibinfo {pages} {073008} (\bibinfo {year}
  {2009})}\BibitemShut {NoStop}%
\bibitem [{\citenamefont {Deffner}\ and\ \citenamefont
  {Lutz}(2011)}]{deffner2011nonequilibrium}%
  \BibitemOpen
  \bibfield  {author} {\bibinfo {author} {\bibfnamefont {S.}~\bibnamefont
  {Deffner}}\ and\ \bibinfo {author} {\bibfnamefont {E.}~\bibnamefont {Lutz}},\
  }\enquote {\bibinfo {title} {Nonequilibrium Entropy Production for Open
  Quantum Systems},}\ \href {\doibase 10.1103/PhysRevLett.107.140404}
  {\bibfield  {journal} {\bibinfo  {journal} {Phys. Rev. Lett.}\ }\textbf
  {\bibinfo {volume} {107}},\ \bibinfo {pages} {140404} (\bibinfo {year}
  {2011})}\BibitemShut {NoStop}%
\bibitem [{\citenamefont {Boukobza}\ and\ \citenamefont
  {Ritsch}(2013)}]{boukobza2013breaking}%
  \BibitemOpen
  \bibfield  {author} {\bibinfo {author} {\bibfnamefont {E.}~\bibnamefont
  {Boukobza}}\ and\ \bibinfo {author} {\bibfnamefont {H.}~\bibnamefont
  {Ritsch}},\ }\enquote {\bibinfo {title} {Breaking the Carnot limit without
  violating the second law: A thermodynamic analysis of off-resonant quantum
  light generation},}\ \href {\doibase 10.1103/PhysRevA.87.063845} {\bibfield
  {journal} {\bibinfo  {journal} {Phys. Rev. A}\ }\textbf {\bibinfo {volume}
  {87}},\ \bibinfo {pages} {063845} (\bibinfo {year} {2013})}\BibitemShut
  {NoStop}%
\bibitem [{\citenamefont {Sagawa}(2013)}]{sagawa2013second}%
  \BibitemOpen
  \bibfield  {author} {\bibinfo {author} {\bibfnamefont {T.}~\bibnamefont
  {Sagawa}},\ }in\ \href {\doibase 10.1142/9789814425193_0003}
  {{\selectlanguage {English}\emph {\bibinfo {booktitle} {Lectures on Quantum
  Computing, Thermodynamics and Statistical Physics}}}},\ \bibinfo {editor}
  {edited by\ \bibinfo {editor} {\bibfnamefont {M.}~\bibnamefont {Nakahara}}\
  and\ \bibinfo {editor} {\bibfnamefont {S.}~\bibnamefont {Tanaka}}}\ (\bibinfo
   {publisher} {World Scientific},\ \bibinfo {address} {Singapore},\ \bibinfo
  {year} {2013})\ pp.\ \bibinfo {pages} {125--190}\BibitemShut {NoStop}%
\bibitem [{\citenamefont {Argentieri}\ \emph {et~al.}(2014)\citenamefont
  {Argentieri}, \citenamefont {Benatti}, \citenamefont {Floreanini},\ and\
  \citenamefont {Pezzutto}}]{argentieri2014violation}%
  \BibitemOpen
  \bibfield  {author} {\bibinfo {author} {\bibfnamefont {G.}~\bibnamefont
  {Argentieri}}, \bibinfo {author} {\bibfnamefont {F.}~\bibnamefont {Benatti}},
  \bibinfo {author} {\bibfnamefont {R.}~\bibnamefont {Floreanini}}, \ and\
  \bibinfo {author} {\bibfnamefont {M.}~\bibnamefont {Pezzutto}},\ }\enquote
  {\bibinfo {title} {Violations of the second law of thermodynamics by a
  non-completely positive dynamics},}\ \href {\doibase
  10.1209/0295-5075/107/50007} {\bibfield  {journal} {\bibinfo  {journal} {EPL
  (Europhys. Lett.)}\ }\textbf {\bibinfo {volume} {107}},\ \bibinfo {pages}
  {50007} (\bibinfo {year} {2014})}\BibitemShut {NoStop}%
\bibitem [{\citenamefont {Binder}\ \emph
  {et~al.}(2015{\natexlab{a}})\citenamefont {Binder}, \citenamefont
  {Vinjanampathy}, \citenamefont {Modi},\ and\ \citenamefont
  {Goold}}]{binder2015quantum}%
  \BibitemOpen
  \bibfield  {author} {\bibinfo {author} {\bibfnamefont {F.}~\bibnamefont
  {Binder}}, \bibinfo {author} {\bibfnamefont {S.}~\bibnamefont
  {Vinjanampathy}}, \bibinfo {author} {\bibfnamefont {K.}~\bibnamefont {Modi}},
  \ and\ \bibinfo {author} {\bibfnamefont {J.}~\bibnamefont {Goold}},\
  }\enquote {\bibinfo {title} {Quantum thermodynamics of general quantum
  processes},}\ \href {\doibase 10.1103/PhysRevE.91.032119} {\bibfield
  {journal} {\bibinfo  {journal} {Phys. Rev. E}\ }\textbf {\bibinfo {volume}
  {91}},\ \bibinfo {pages} {032119} (\bibinfo {year}
  {2015}{\natexlab{a}})}\BibitemShut {NoStop}%
\bibitem [{\citenamefont {Brandner}\ and\ \citenamefont
  {Seifert}(2016)}]{brandner2016periodic}%
  \BibitemOpen
  \bibfield  {author} {\bibinfo {author} {\bibfnamefont {K.}~\bibnamefont
  {Brandner}}\ and\ \bibinfo {author} {\bibfnamefont {U.}~\bibnamefont
  {Seifert}},\ }\enquote {\bibinfo {title} {Periodic thermodynamics of open
  quantum systems},}\ \href {\doibase 10.1103/PhysRevE.93.062134} {\bibfield
  {journal} {\bibinfo  {journal} {Phys. Rev. E}\ }\textbf {\bibinfo {volume}
  {93}},\ \bibinfo {pages} {062134} (\bibinfo {year} {2016})}\BibitemShut
  {NoStop}%
\bibitem [{\citenamefont {Breuer}\ and\ \citenamefont
  {Petruccione}(2002)}]{breuerbook}%
  \BibitemOpen
  \bibfield  {author} {\bibinfo {author} {\bibfnamefont {H.-P.}\ \bibnamefont
  {Breuer}}\ and\ \bibinfo {author} {\bibfnamefont {F.}~\bibnamefont
  {Petruccione}},\ }\href@noop {} {\emph {\bibinfo {title} {The Theory of Open
  Quantum Systems}}}\ (\bibinfo  {publisher} {Oxford University Press},\
  \bibinfo {year} {2002})\BibitemShut {NoStop}%
\bibitem [{\citenamefont {Spohn}(1978)}]{spohn1978entropy}%
  \BibitemOpen
  \bibfield  {author} {\bibinfo {author} {\bibfnamefont {H.}~\bibnamefont
  {Spohn}},\ }\enquote {\bibinfo {title} {Entropy production for quantum
  dynamical semigroups},}\ \href {\doibase 10.1063/1.523789} {\bibfield
  {journal} {\bibinfo  {journal} {J. Math. Phys.}\ }\textbf {\bibinfo {volume}
  {19}},\ \bibinfo {pages} {1227} (\bibinfo {year} {1978})}\BibitemShut
  {NoStop}%
\bibitem [{\citenamefont {Anders}\ and\ \citenamefont
  {Giovannetti}(2013)}]{anders2013thermodynamics}%
  \BibitemOpen
  \bibfield  {author} {\bibinfo {author} {\bibfnamefont {J.}~\bibnamefont
  {Anders}}\ and\ \bibinfo {author} {\bibfnamefont {V.}~\bibnamefont
  {Giovannetti}},\ }\enquote {\bibinfo {title} {Thermodynamics of discrete
  quantum processes},}\ \href {\doibase 10.1088/1367-2630/15/3/033022}
  {\bibfield  {journal} {\bibinfo  {journal} {New J. Phys.}\ }\textbf {\bibinfo
  {volume} {15}},\ \bibinfo {pages} {033022} (\bibinfo {year}
  {2013})}\BibitemShut {NoStop}%
\bibitem [{\citenamefont {Alicki}\ and\ \citenamefont
  {Fannes}(2013)}]{alicki2013entanglement}%
  \BibitemOpen
  \bibfield  {author} {\bibinfo {author} {\bibfnamefont {R.}~\bibnamefont
  {Alicki}}\ and\ \bibinfo {author} {\bibfnamefont {M.}~\bibnamefont
  {Fannes}},\ }\enquote {\bibinfo {title} {Entanglement boost for extractable
  work from ensembles of quantum batteries},}\ \href {\doibase
  10.1103/PhysRevE.87.042123} {\bibfield  {journal} {\bibinfo  {journal} {Phys.
  Rev. E}\ }\textbf {\bibinfo {volume} {87}},\ \bibinfo {pages} {042123}
  (\bibinfo {year} {2013})}\BibitemShut {NoStop}%
\bibitem [{\citenamefont {Gelbwaser-Klimovsky}\ \emph
  {et~al.}(2013{\natexlab{a}})\citenamefont {Gelbwaser-Klimovsky},
  \citenamefont {Alicki},\ and\ \citenamefont {Kurizki}}]{gelbwaser2013work}%
  \BibitemOpen
  \bibfield  {author} {\bibinfo {author} {\bibfnamefont {D.}~\bibnamefont
  {Gelbwaser-Klimovsky}}, \bibinfo {author} {\bibfnamefont {R.}~\bibnamefont
  {Alicki}}, \ and\ \bibinfo {author} {\bibfnamefont {G.}~\bibnamefont
  {Kurizki}},\ }\enquote {\bibinfo {title} {Work and energy gain of heat-pumped
  quantized amplifiers},}\ \href {\doibase 10.1209/0295-5075/103/60005}
  {\bibfield  {journal} {\bibinfo  {journal} {EPL (Europhys. Lett.)}\ }\textbf
  {\bibinfo {volume} {103}},\ \bibinfo {pages} {60005} (\bibinfo {year}
  {2013}{\natexlab{a}})}\BibitemShut {NoStop}%
\bibitem [{\citenamefont {Hovhannisyan}\ \emph {et~al.}(2013)\citenamefont
  {Hovhannisyan}, \citenamefont {Perarnau-Llobet}, \citenamefont {Huber},\ and\
  \citenamefont {Ac\'{\i}n}}]{hovhannisyan2013entanglement}%
  \BibitemOpen
  \bibfield  {author} {\bibinfo {author} {\bibfnamefont {K.~V.}\ \bibnamefont
  {Hovhannisyan}}, \bibinfo {author} {\bibfnamefont {M.}~\bibnamefont
  {Perarnau-Llobet}}, \bibinfo {author} {\bibfnamefont {M.}~\bibnamefont
  {Huber}}, \ and\ \bibinfo {author} {\bibfnamefont {A.}~\bibnamefont
  {Ac\'{\i}n}},\ }\enquote {\bibinfo {title} {Entanglement Generation is Not
  Necessary for Optimal Work Extraction},}\ \href {\doibase
  10.1103/PhysRevLett.111.240401} {\bibfield  {journal} {\bibinfo  {journal}
  {Phys. Rev. Lett.}\ }\textbf {\bibinfo {volume} {111}},\ \bibinfo {pages}
  {240401} (\bibinfo {year} {2013})}\BibitemShut {NoStop}%
\bibitem [{\citenamefont {Binder}\ \emph
  {et~al.}(2015{\natexlab{b}})\citenamefont {Binder}, \citenamefont
  {Vinjanampathy}, \citenamefont {Modi},\ and\ \citenamefont
  {Goold}}]{binder2015quantacell}%
  \BibitemOpen
  \bibfield  {author} {\bibinfo {author} {\bibfnamefont {F.~C.}\ \bibnamefont
  {Binder}}, \bibinfo {author} {\bibfnamefont {S.}~\bibnamefont
  {Vinjanampathy}}, \bibinfo {author} {\bibfnamefont {K.}~\bibnamefont {Modi}},
  \ and\ \bibinfo {author} {\bibfnamefont {J.}~\bibnamefont {Goold}},\
  }\enquote {\bibinfo {title} {Quantacell: powerful charging of quantum
  batteries},}\ \href {\doibase 10.1088/1367-2630/17/7/075015} {\bibfield
  {journal} {\bibinfo  {journal} {New J. Phys.}\ }\textbf {\bibinfo {volume}
  {17}},\ \bibinfo {pages} {075015} (\bibinfo {year}
  {2015}{\natexlab{b}})}\BibitemShut {NoStop}%
\bibitem [{\citenamefont {Perarnau-Llobet}\ \emph {et~al.}(2015)\citenamefont
  {Perarnau-Llobet}, \citenamefont {Hovhannisyan}, \citenamefont {Huber},
  \citenamefont {Skrzypczyk}, \citenamefont {Brunner},\ and\ \citenamefont
  {Ac\'{\i}n}}]{perarnau2015extractable}%
  \BibitemOpen
  \bibfield  {author} {\bibinfo {author} {\bibfnamefont {M.}~\bibnamefont
  {Perarnau-Llobet}}, \bibinfo {author} {\bibfnamefont {K.~V.}\ \bibnamefont
  {Hovhannisyan}}, \bibinfo {author} {\bibfnamefont {M.}~\bibnamefont {Huber}},
  \bibinfo {author} {\bibfnamefont {P.}~\bibnamefont {Skrzypczyk}}, \bibinfo
  {author} {\bibfnamefont {N.}~\bibnamefont {Brunner}}, \ and\ \bibinfo
  {author} {\bibfnamefont {A.}~\bibnamefont {Ac\'{\i}n}},\ }\enquote {\bibinfo
  {title} {Extractable Work from Correlations},}\ \href {\doibase
  10.1103/PhysRevX.5.041011} {\bibfield  {journal} {\bibinfo  {journal} {Phys.
  Rev. X}\ }\textbf {\bibinfo {volume} {5}},\ \bibinfo {pages} {041011}
  (\bibinfo {year} {2015})}\BibitemShut {NoStop}%
\bibitem [{\citenamefont {Skrzypczyk}\ \emph {et~al.}(2015)\citenamefont
  {Skrzypczyk}, \citenamefont {Silva},\ and\ \citenamefont
  {Brunner}}]{skrzypczyk2015passivity}%
  \BibitemOpen
  \bibfield  {author} {\bibinfo {author} {\bibfnamefont {P.}~\bibnamefont
  {Skrzypczyk}}, \bibinfo {author} {\bibfnamefont {R.}~\bibnamefont {Silva}}, \
  and\ \bibinfo {author} {\bibfnamefont {N.}~\bibnamefont {Brunner}},\
  }\enquote {\bibinfo {title} {Passivity, complete passivity, and virtual
  temperatures},}\ \href {\doibase 10.1103/PhysRevE.91.052133} {\bibfield
  {journal} {\bibinfo  {journal} {Phys. Rev. E}\ }\textbf {\bibinfo {volume}
  {91}},\ \bibinfo {pages} {052133} (\bibinfo {year} {2015})}\BibitemShut
  {NoStop}%
\bibitem [{\citenamefont {Brown}\ \emph {et~al.}(2016)\citenamefont {Brown},
  \citenamefont {Friis},\ and\ \citenamefont {Huber}}]{brown2016passivity}%
  \BibitemOpen
  \bibfield  {author} {\bibinfo {author} {\bibfnamefont {E.~G.}\ \bibnamefont
  {Brown}}, \bibinfo {author} {\bibfnamefont {N.}~\bibnamefont {Friis}}, \ and\
  \bibinfo {author} {\bibfnamefont {M.}~\bibnamefont {Huber}},\ }\enquote
  {\bibinfo {title} {Passivity and practical work extraction using Gaussian
  operations},}\ \href {\doibase 10.1088/1367-2630/18/11/113028} {\bibfield
  {journal} {\bibinfo  {journal} {New J. Phys.}\ }\textbf {\bibinfo {volume}
  {18}},\ \bibinfo {pages} {113028} (\bibinfo {year} {2016})}\BibitemShut
  {NoStop}%
\bibitem [{\citenamefont {De~Palma}\ \emph {et~al.}(2016)\citenamefont
  {De~Palma}, \citenamefont {Mari}, \citenamefont {Lloyd},\ and\ \citenamefont
  {Giovannetti}}]{depalma2016passive}%
  \BibitemOpen
  \bibfield  {author} {\bibinfo {author} {\bibfnamefont {G.}~\bibnamefont
  {De~Palma}}, \bibinfo {author} {\bibfnamefont {A.}~\bibnamefont {Mari}},
  \bibinfo {author} {\bibfnamefont {S.}~\bibnamefont {Lloyd}}, \ and\ \bibinfo
  {author} {\bibfnamefont {V.}~\bibnamefont {Giovannetti}},\ }\enquote
  {\bibinfo {title} {Passive states as optimal inputs for single-jump lossy
  quantum channels},}\ \href {\doibase 10.1103/PhysRevA.93.062328} {\bibfield
  {journal} {\bibinfo  {journal} {Phys. Rev. A}\ }\textbf {\bibinfo {volume}
  {93}},\ \bibinfo {pages} {062328} (\bibinfo {year} {2016})}\BibitemShut
  {NoStop}%
\bibitem [{\citenamefont {Bruschi}(2017)}]{bruschi2017gravitational}%
  \BibitemOpen
  \bibfield  {author} {\bibinfo {author} {\bibfnamefont {D.~E.}\ \bibnamefont
  {Bruschi}},\ }\enquote {\bibinfo {title} {On the gravitational nature of
  energy},}\ \href {https://arxiv.org/abs/1701.00699} {\bibfield  {journal}
  {\bibinfo  {journal} {arXiv preprint arXiv:1701.00699}\ } (\bibinfo {year}
  {2017})}\BibitemShut {NoStop}%
\bibitem [{\citenamefont {Levy}\ \emph {et~al.}(2016)\citenamefont {Levy},
  \citenamefont {Di\'osi},\ and\ \citenamefont {Kosloff}}]{levy2016quantum}%
  \BibitemOpen
  \bibfield  {author} {\bibinfo {author} {\bibfnamefont {A.}~\bibnamefont
  {Levy}}, \bibinfo {author} {\bibfnamefont {L.}~\bibnamefont {Di\'osi}}, \
  and\ \bibinfo {author} {\bibfnamefont {R.}~\bibnamefont {Kosloff}},\
  }\enquote {\bibinfo {title} {Quantum flywheel},}\ \href {\doibase
  10.1103/PhysRevA.93.052119} {\bibfield  {journal} {\bibinfo  {journal} {Phys.
  Rev. A}\ }\textbf {\bibinfo {volume} {93}},\ \bibinfo {pages} {052119}
  (\bibinfo {year} {2016})}\BibitemShut {NoStop}%
\bibitem [{\citenamefont {Mari}\ \emph {et~al.}(2014)\citenamefont {Mari},
  \citenamefont {Giovannetti},\ and\ \citenamefont {Holevo}}]{mari2014quantum}%
  \BibitemOpen
  \bibfield  {author} {\bibinfo {author} {\bibfnamefont {A.}~\bibnamefont
  {Mari}}, \bibinfo {author} {\bibfnamefont {V.}~\bibnamefont {Giovannetti}}, \
  and\ \bibinfo {author} {\bibfnamefont {A.~S.}\ \bibnamefont {Holevo}},\
  }\enquote {\bibinfo {title} {Quantum state majorization at the output of
  bosonic Gaussian channels},}\ \href {\doibase 10.1038/ncomms4826} {\bibfield
  {journal} {\bibinfo  {journal} {Nat. Commun.}\ }\textbf {\bibinfo {volume}
  {5}},\ \bibinfo {pages} {3826} (\bibinfo {year} {2014})}\BibitemShut
  {NoStop}%
\bibitem [{\citenamefont {Gardiner}\ and\ \citenamefont
  {Zoller}(2000)}]{gardinerbook}%
  \BibitemOpen
  \bibfield  {author} {\bibinfo {author} {\bibfnamefont {C.~W.}\ \bibnamefont
  {Gardiner}}\ and\ \bibinfo {author} {\bibfnamefont {P.}~\bibnamefont
  {Zoller}},\ }\href@noop {} {\emph {\bibinfo {title} {Quantum Noise}}},\
  \bibinfo {edition} {2nd}\ ed.\ (\bibinfo  {publisher} {Springer-Verlag},\
  \bibinfo {address} {Berlin},\ \bibinfo {year} {2000})\BibitemShut {NoStop}%
\bibitem [{\citenamefont {Alipour}\ \emph {et~al.}(2016)\citenamefont
  {Alipour}, \citenamefont {Benatti}, \citenamefont {Bakhshinezhad},
  \citenamefont {Afsary}, \citenamefont {Marcantoni},\ and\ \citenamefont
  {Rezakhani}}]{alipour2016correlations}%
  \BibitemOpen
  \bibfield  {author} {\bibinfo {author} {\bibfnamefont {S.}~\bibnamefont
  {Alipour}}, \bibinfo {author} {\bibfnamefont {F.}~\bibnamefont {Benatti}},
  \bibinfo {author} {\bibfnamefont {F.}~\bibnamefont {Bakhshinezhad}}, \bibinfo
  {author} {\bibfnamefont {M.}~\bibnamefont {Afsary}}, \bibinfo {author}
  {\bibfnamefont {S.}~\bibnamefont {Marcantoni}}, \ and\ \bibinfo {author}
  {\bibfnamefont {A.~T.}\ \bibnamefont {Rezakhani}},\ }\enquote {\bibinfo
  {title} {Correlations in quantum thermodynamics: Heat, work, and entropy
  production},}\ \href {\doibase 10.1038/srep35568} {\bibfield  {journal}
  {\bibinfo  {journal} {Sci. Rep.}\ }\textbf {\bibinfo {volume} {6}},\ \bibinfo
  {pages} {35568} (\bibinfo {year} {2016})}\BibitemShut {NoStop}%
\bibitem [{\citenamefont {Schl{\"o}gl}(1966)}]{schloegl1966zur}%
  \BibitemOpen
  \bibfield  {author} {\bibinfo {author} {\bibfnamefont {F.}~\bibnamefont
  {Schl{\"o}gl}},\ }\enquote {\bibinfo {title} {Zur statistischen Theorie der
  Entropieproduktion in nicht abgeschlossenen Systemen},}\ \href {\doibase
  10.1007/BF01362471} {\bibfield  {journal} {\bibinfo  {journal} {Z. Phys.}\
  }\textbf {\bibinfo {volume} {191}},\ \bibinfo {pages} {81} (\bibinfo {year}
  {1966})}\BibitemShut {NoStop}%
\bibitem [{\citenamefont {Ekert}\ and\ \citenamefont
  {Knight}(1990)}]{ekert1990canonical}%
  \BibitemOpen
  \bibfield  {author} {\bibinfo {author} {\bibfnamefont {A.~K.}\ \bibnamefont
  {Ekert}}\ and\ \bibinfo {author} {\bibfnamefont {P.~L.}\ \bibnamefont
  {Knight}},\ }\enquote {\bibinfo {title} {Canonical transformation and decay
  into phase-sensitive reservoirs},}\ \href {\doibase 10.1103/PhysRevA.42.487}
  {\bibfield  {journal} {\bibinfo  {journal} {Phys. Rev. A}\ }\textbf {\bibinfo
  {volume} {42}},\ \bibinfo {pages} {487} (\bibinfo {year} {1990})}\BibitemShut
  {NoStop}%
\bibitem [{\citenamefont {Gelbwaser-Klimovsky}\ \emph
  {et~al.}(2013{\natexlab{b}})\citenamefont {Gelbwaser-Klimovsky},
  \citenamefont {Alicki},\ and\ \citenamefont
  {Kurizki}}]{gelbwaser2013minimal}%
  \BibitemOpen
  \bibfield  {author} {\bibinfo {author} {\bibfnamefont {D.}~\bibnamefont
  {Gelbwaser-Klimovsky}}, \bibinfo {author} {\bibfnamefont {R.}~\bibnamefont
  {Alicki}}, \ and\ \bibinfo {author} {\bibfnamefont {G.}~\bibnamefont
  {Kurizki}},\ }\enquote {\bibinfo {title} {Minimal universal quantum heat
  machine},}\ \href {\doibase 10.1103/PhysRevE.87.012140} {\bibfield  {journal}
  {\bibinfo  {journal} {Phys. Rev. E}\ }\textbf {\bibinfo {volume} {87}},\
  \bibinfo {pages} {012140} (\bibinfo {year} {2013}{\natexlab{b}})}\BibitemShut
  {NoStop}%
\bibitem [{\citenamefont {Mukherjee}\ \emph {et~al.}(2016)\citenamefont
  {Mukherjee}, \citenamefont {Niedenzu}, \citenamefont {Kofman},\ and\
  \citenamefont {Kurizki}}]{mukherjee2016speed}%
  \BibitemOpen
  \bibfield  {author} {\bibinfo {author} {\bibfnamefont {V.}~\bibnamefont
  {Mukherjee}}, \bibinfo {author} {\bibfnamefont {W.}~\bibnamefont {Niedenzu}},
  \bibinfo {author} {\bibfnamefont {A.~G.}\ \bibnamefont {Kofman}}, \ and\
  \bibinfo {author} {\bibfnamefont {G.}~\bibnamefont {Kurizki}},\ }\enquote
  {\bibinfo {title} {Speed and efficiency limits of multilevel incoherent heat
  engines},}\ \href {\doibase 10.1103/PhysRevE.94.062109} {\bibfield  {journal}
  {\bibinfo  {journal} {Phys. Rev. E}\ }\textbf {\bibinfo {volume} {94}},\
  \bibinfo {pages} {062109} (\bibinfo {year} {2016})}\BibitemShut {NoStop}%
\bibitem [{\citenamefont {Erker}\ \emph {et~al.}(2017)\citenamefont {Erker},
  \citenamefont {Mitchison}, \citenamefont {Silva}, \citenamefont {Woods},
  \citenamefont {Brunner},\ and\ \citenamefont {Huber}}]{erker2017autonomous}%
  \BibitemOpen
  \bibfield  {author} {\bibinfo {author} {\bibfnamefont {P.}~\bibnamefont
  {Erker}}, \bibinfo {author} {\bibfnamefont {M.~T.}\ \bibnamefont
  {Mitchison}}, \bibinfo {author} {\bibfnamefont {R.}~\bibnamefont {Silva}},
  \bibinfo {author} {\bibfnamefont {M.~P.}\ \bibnamefont {Woods}}, \bibinfo
  {author} {\bibfnamefont {N.}~\bibnamefont {Brunner}}, \ and\ \bibinfo
  {author} {\bibfnamefont {M.}~\bibnamefont {Huber}},\ }\enquote {\bibinfo
  {title} {Autonomous Quantum Clocks: Does Thermodynamics Limit Our Ability to
  Measure Time?}}\ \href {\doibase 10.1103/PhysRevX.7.031022} {\bibfield
  {journal} {\bibinfo  {journal} {Phys. Rev. X}\ }\textbf {\bibinfo {volume}
  {7}},\ \bibinfo {pages} {031022} (\bibinfo {year} {2017})}\BibitemShut
  {NoStop}%
\bibitem [{\citenamefont {Woods}\ \emph {et~al.}(2016)\citenamefont {Woods},
  \citenamefont {Silva},\ and\ \citenamefont
  {Oppenheim}}]{woods2016autonomous}%
  \BibitemOpen
  \bibfield  {author} {\bibinfo {author} {\bibfnamefont {M.~P.}\ \bibnamefont
  {Woods}}, \bibinfo {author} {\bibfnamefont {R.}~\bibnamefont {Silva}}, \ and\
  \bibinfo {author} {\bibfnamefont {J.}~\bibnamefont {Oppenheim}},\ }\enquote
  {\bibinfo {title} {Autonomous quantum machines and finite sized clocks},}\
  \href {https://arxiv.org/abs/1607.04591} {\bibfield  {journal} {\bibinfo
  {journal} {arXiv preprint arXiv:1607.04591}\ } (\bibinfo {year}
  {2016})}\BibitemShut {NoStop}%
\bibitem [{\citenamefont {Graham}(1987)}]{graham1987squeezing}%
  \BibitemOpen
  \bibfield  {author} {\bibinfo {author} {\bibfnamefont {R.}~\bibnamefont
  {Graham}},\ }\enquote {\bibinfo {title} {Squeezing and Frequency Changes in
  Harmonic Oscillations},}\ \href {\doibase 10.1080/09500348714550801}
  {\bibfield  {journal} {\bibinfo  {journal} {J. Mod. Opt.}\ }\textbf {\bibinfo
  {volume} {34}},\ \bibinfo {pages} {873} (\bibinfo {year} {1987})}\BibitemShut
  {NoStop}%
\bibitem [{\citenamefont {Agarwal}\ and\ \citenamefont
  {Kumar}(1991)}]{agarwal1991exact}%
  \BibitemOpen
  \bibfield  {author} {\bibinfo {author} {\bibfnamefont {G.~S.}\ \bibnamefont
  {Agarwal}}\ and\ \bibinfo {author} {\bibfnamefont {S.~A.}\ \bibnamefont
  {Kumar}},\ }\enquote {\bibinfo {title} {Exact quantum-statistical dynamics of
  an oscillator with time-dependent frequency and generation of nonclassical
  states},}\ \href {\doibase 10.1103/PhysRevLett.67.3665} {\bibfield  {journal}
  {\bibinfo  {journal} {Phys. Rev. Lett.}\ }\textbf {\bibinfo {volume} {67}},\
  \bibinfo {pages} {3665} (\bibinfo {year} {1991})}\BibitemShut {NoStop}%
\bibitem [{\citenamefont {Averbukh}\ \emph {et~al.}(1994)\citenamefont
  {Averbukh}, \citenamefont {Sherman},\ and\ \citenamefont
  {Kurizki}}]{averbukh1994enhanced}%
  \BibitemOpen
  \bibfield  {author} {\bibinfo {author} {\bibfnamefont {I.}~\bibnamefont
  {Averbukh}}, \bibinfo {author} {\bibfnamefont {B.}~\bibnamefont {Sherman}}, \
  and\ \bibinfo {author} {\bibfnamefont {G.}~\bibnamefont {Kurizki}},\
  }\enquote {\bibinfo {title} {Enhanced squeezing by periodic frequency
  modulation under parametric instability conditions},}\ \href {\doibase
  10.1103/PhysRevA.50.5301} {\bibfield  {journal} {\bibinfo  {journal} {Phys.
  Rev. A}\ }\textbf {\bibinfo {volume} {50}},\ \bibinfo {pages} {5301}
  (\bibinfo {year} {1994})}\BibitemShut {NoStop}%
\bibitem [{\citenamefont {Geva}\ and\ \citenamefont
  {Kosloff}(1992)}]{geva1992quantum}%
  \BibitemOpen
  \bibfield  {author} {\bibinfo {author} {\bibfnamefont {E.}~\bibnamefont
  {Geva}}\ and\ \bibinfo {author} {\bibfnamefont {R.}~\bibnamefont {Kosloff}},\
  }\enquote {\bibinfo {title} {A quantum-mechanical heat engine operating in
  finite time. A model consisting of spin-1/2 systems as the working fluid},}\
  \href {\doibase 10.1063/1.461951} {\bibfield  {journal} {\bibinfo  {journal}
  {J. Chem. Phys.}\ }\textbf {\bibinfo {volume} {96}},\ \bibinfo {pages} {3054}
  (\bibinfo {year} {1992})}\BibitemShut {NoStop}%
\bibitem [{\citenamefont {Feldmann}\ and\ \citenamefont
  {Kosloff}(2004)}]{feldmann2004characteristics}%
  \BibitemOpen
  \bibfield  {author} {\bibinfo {author} {\bibfnamefont {T.}~\bibnamefont
  {Feldmann}}\ and\ \bibinfo {author} {\bibfnamefont {R.}~\bibnamefont
  {Kosloff}},\ }\enquote {\bibinfo {title} {Characteristics of the limit cycle
  of a reciprocating quantum heat engine},}\ \href {\doibase
  10.1103/PhysRevE.70.046110} {\bibfield  {journal} {\bibinfo  {journal} {Phys.
  Rev. E}\ }\textbf {\bibinfo {volume} {70}},\ \bibinfo {pages} {046110}
  (\bibinfo {year} {2004})}\BibitemShut {NoStop}%
\bibitem [{\citenamefont {Quan}\ \emph {et~al.}(2007)\citenamefont {Quan},
  \citenamefont {Liu}, \citenamefont {Sun},\ and\ \citenamefont
  {Nori}}]{quan2007quantum}%
  \BibitemOpen
  \bibfield  {author} {\bibinfo {author} {\bibfnamefont {H.~T.}\ \bibnamefont
  {Quan}}, \bibinfo {author} {\bibfnamefont {Y.-x.}\ \bibnamefont {Liu}},
  \bibinfo {author} {\bibfnamefont {C.~P.}\ \bibnamefont {Sun}}, \ and\
  \bibinfo {author} {\bibfnamefont {F.}~\bibnamefont {Nori}},\ }\enquote
  {\bibinfo {title} {Quantum thermodynamic cycles and quantum heat engines},}\
  \href {\doibase 10.1103/PhysRevE.76.031105} {\bibfield  {journal} {\bibinfo
  {journal} {Phys. Rev. E}\ }\textbf {\bibinfo {volume} {76}},\ \bibinfo
  {pages} {031105} (\bibinfo {year} {2007})}\BibitemShut {NoStop}%
\bibitem [{\citenamefont {del Campo}\ \emph {et~al.}(2014)\citenamefont {del
  Campo}, \citenamefont {Goold},\ and\ \citenamefont
  {Paternostro}}]{delcampo2014more}%
  \BibitemOpen
  \bibfield  {author} {\bibinfo {author} {\bibfnamefont {A.}~\bibnamefont {del
  Campo}}, \bibinfo {author} {\bibfnamefont {J.}~\bibnamefont {Goold}}, \ and\
  \bibinfo {author} {\bibfnamefont {M.}~\bibnamefont {Paternostro}},\ }\enquote
  {\bibinfo {title} {More bang for your buck: Super-adiabatic quantum
  engines},}\ \href {\doibase 10.1038/srep06208} {\bibfield  {journal}
  {\bibinfo  {journal} {Sci. Rep.}\ }\textbf {\bibinfo {volume} {4}},\ \bibinfo
  {pages} {6208} (\bibinfo {year} {2014})}\BibitemShut {NoStop}%
\bibitem [{\citenamefont {Hatano}\ and\ \citenamefont
  {Sasa}(2001)}]{hatano2001steady}%
  \BibitemOpen
  \bibfield  {author} {\bibinfo {author} {\bibfnamefont {T.}~\bibnamefont
  {Hatano}}\ and\ \bibinfo {author} {\bibfnamefont {S.-i.}\ \bibnamefont
  {Sasa}},\ }\enquote {\bibinfo {title} {Steady-State Thermodynamics of
  Langevin Systems},}\ \href {\doibase 10.1103/PhysRevLett.86.3463} {\bibfield
  {journal} {\bibinfo  {journal} {Phys. Rev. Lett.}\ }\textbf {\bibinfo
  {volume} {86}},\ \bibinfo {pages} {3463} (\bibinfo {year}
  {2001})}\BibitemShut {NoStop}%
\bibitem [{\citenamefont {Gardas}\ and\ \citenamefont
  {Deffner}(2015)}]{gardas2015thermodynamic}%
  \BibitemOpen
  \bibfield  {author} {\bibinfo {author} {\bibfnamefont {B.}~\bibnamefont
  {Gardas}}\ and\ \bibinfo {author} {\bibfnamefont {S.}~\bibnamefont
  {Deffner}},\ }\enquote {\bibinfo {title} {Thermodynamic universality of
  quantum Carnot engines},}\ \href {\doibase 10.1103/PhysRevE.92.042126}
  {\bibfield  {journal} {\bibinfo  {journal} {Phys. Rev. E}\ }\textbf {\bibinfo
  {volume} {92}},\ \bibinfo {pages} {042126} (\bibinfo {year}
  {2015})}\BibitemShut {NoStop}%
\bibitem [{\citenamefont {Misra}\ \emph {et~al.}(2015)\citenamefont {Misra},
  \citenamefont {Singh}, \citenamefont {Bera},\ and\ \citenamefont
  {Rajagopal}}]{misra2015quantum}%
  \BibitemOpen
  \bibfield  {author} {\bibinfo {author} {\bibfnamefont {A.}~\bibnamefont
  {Misra}}, \bibinfo {author} {\bibfnamefont {U.}~\bibnamefont {Singh}},
  \bibinfo {author} {\bibfnamefont {M.~N.}\ \bibnamefont {Bera}}, \ and\
  \bibinfo {author} {\bibfnamefont {A.~K.}\ \bibnamefont {Rajagopal}},\
  }\enquote {\bibinfo {title} {Quantum R\'enyi relative entropies affirm
  universality of thermodynamics},}\ \href {\doibase
  10.1103/PhysRevE.92.042161} {\bibfield  {journal} {\bibinfo  {journal} {Phys.
  Rev. E}\ }\textbf {\bibinfo {volume} {92}},\ \bibinfo {pages} {042161}
  (\bibinfo {year} {2015})}\BibitemShut {NoStop}%
\bibitem [{\citenamefont {Qi}\ \emph {et~al.}(2016)\citenamefont {Qi},
  \citenamefont {Wilde},\ and\ \citenamefont {Guha}}]{qi2016thermal}%
  \BibitemOpen
  \bibfield  {author} {\bibinfo {author} {\bibfnamefont {H.}~\bibnamefont
  {Qi}}, \bibinfo {author} {\bibfnamefont {M.~M.}\ \bibnamefont {Wilde}}, \
  and\ \bibinfo {author} {\bibfnamefont {S.}~\bibnamefont {Guha}},\ }\enquote
  {\bibinfo {title} {Thermal states minimize the output entropy of single-mode
  phase-insensitive Gaussian channels with an input entropy constraint},}\
  \href {https://arxiv.org/abs/1607.05262} {\bibfield  {journal} {\bibinfo
  {journal} {arXiv preprint arXiv:1607.05262}\ } (\bibinfo {year}
  {2016})}\BibitemShut {NoStop}%
\end{thebibliography}
\end{document}